\documentclass[
aps,%
12pt,%
final,%
notitlepage,%
oneside,%
onecolumn,%
nobibnotes,%
nofootinbib,%
groupedaddress,%
noshowpacs,%
centertags]%
{revtex4}
\usepackage [cp866]{inputenc}
\usepackage [english]{babel}

\begin{document}
\selectlanguage{english}

\title{Analysis of the triplet production  \\
by the circularly polarized photon at high energies}

\author{\firstname{G.~I.}~\surname{Gakh}}
\author{\firstname{M.~I.}~\surname{Konchatnij}}
\author{\firstname{I.~S.}~\surname{Levandovsky}}
\author{\firstname{N.~P.}~\surname{Merenkov}}
\email{merenkov@kipt.kharkov.ua}
\affiliation{Kharkov Institute of Physics and Technology \\
61108, Akademicheskaya, 1, Kharkov, Ukraine}%

\begin{abstract}
The possibility in principle of the determining high energy photon
circular polarization by the measurement of the created electron
polarization in the process of triplet photoproduction $\gamma
+e^-\rightarrow e^+e^- +e^-$ is investigated. The respective event
number which depend on polarization states of photon and created
electron does not decrease with the growth of the photon energy,
and this circumstance can ensure the high efficiency in such kind
of experiments. We study different double and single distributions
of the created electron (or positron), which allow to probe the
photon circular polarization and to measure its magnitude (the
Stock's parameter $\xi_2$), using the technique of the Sudakov's
variables. Some experimental setups with different rules for event
selection are studied and corresponding numerical estimations are
presented.
\end{abstract}

\maketitle

\section{Introduction}
 It is well known that process of the triplet
production
\begin{equation}\label{1}
 \gamma(k)+e^-(p)\rightarrow e^-(k_1)+e^+(k_2)+e^-(p_1)
\end{equation}
by the high-energy photons on the atomic electrons can be used to
measure the photon linear polarization degree
\cite{BP71,VK73,VM75}. This possibility arises due to azimuthal
asymmetry of the corresponding cross-section, i.e., due to its
dependence on the angle between the plane in which the photon is
polarized, and the plane $({\bf{k}},{\bf{p}}_1)$ where the recoil
electron 3-momentum lies. The detailed description of the
different differential distributions, such as the dependence on
the momentum value, on the polar angle and minimal recorded
momentum of the recoil electron, dependence on the invariant mass
of the created electron-positron pair, on the positron energy and
others, has been investigated in Ref. \cite{BVMP94}. This
single-spin effect is the basis for theoretical background of
polarimeters where the different angular and energy distributions
are used \cite{EKH89}.

The exact expressions for differential and partly integrated cross
sections of the process (1) is very cumbersome and exist in the
complete form only for unpolarized case \cite{Haug85}. At high
collision energy only two (from eight) diagrams contribute with
leading accuracy (neglecting terms of the order of $m^2/s,\, \
s=2(kp), \ m$ is the electron mass) and the corresponding
expressions are essentially simplified. These diagrams (the
so-called Borselino diagrams \cite{B47}) are shown in Fig.1.
Nevertheless, at the boundaries of the final particle phase space
the non-leading terms can be reinforced, and in Ref. \cite{AAK00}
some of such effects had been investigated for the case of
linearly polarized photons.

As regards the photon circular polarization, it can be probed by
at least double-spin effects. In the region of small and
intermediate photon energies the circular polarization can be
measured using double-spin correlation in the Compton scattering.
For example, in Ref. \cite{APS01} the corresponding possibility
was considered for the Compton cross-section asymmetry in the
scattering of photon on polarized electrons. In principle, one can
also measure the polarization of the recoil electron. The
double-spin effects may be used to create polarized electron beams
using the laser photons \cite{KST03}.

At high energies of the photon beams the use of the Compton
scattering is not effective because the Compton cross-section
decreases very fast with the growth of the photon energy. If the
photon energy is large, the cross-section of the electron-positron
pair production, which does not decrease with the growth of the
energy, has become larger than the Compton scattering one. To
estimate the respective energy one can use the asymptotic formulas
for the total cross-sections \cite{AB69}
\begin{equation}\label{2}
\sigma_C\approx\frac{2\pi r_0^2}{x}\ln x\,,\ \
\sigma_{pair}\approx\frac{28 \alpha r_0^2}{9}\ln x\,,
\end{equation}
$$x=\frac{s}{m^2}\,,  \
\alpha=\frac{1}{137}\,,$$ where $r_0=\alpha/m$ is the classical
radius of electron. In the rest system of the initial electron
$(s=2\omega m)$ the photon energy $\omega$ has to be larger than
for about $80\,MeV.$ Thus, to measure the circular polarization of
the photons with the energies more than $100\,MeV$ it is
advantageously to use the process (1) rather than the Compton
scattering.

\vspace{0.6cm}

\begin{minipage}{150 mm}
\begin{center}
\includegraphics[width=0.6\textwidth]{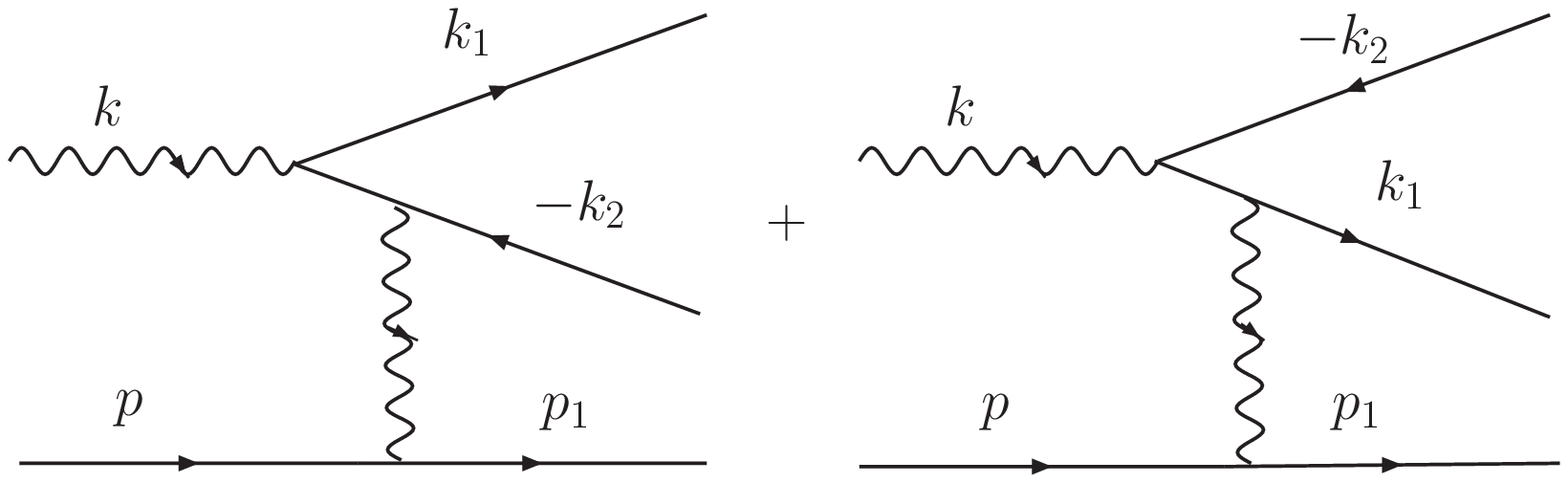}\\
 \vspace{0.2cm} \emph{\textbf{Fig.1.}} {\emph{Borselino diagrams
which give the nondecreasing contribution in the cross section at
high energies and small momentum transferred. }}\\
\end{center}
\end{minipage}

\vspace{0.5cm}

The above estimate of the pair production cross-section is made
taking into account only the  Borselino diagrams. The events,
described by these diagrams, have very specific kinematics in the
rest system of the initial electron, namely: the recoil electron
has small 3-momentum (of the order of $m$) whereas the created
electron-positron pair carries out all the photon energy and moves
along the photon momentum direction. In the reaction c.m.s the
scattered (recoil) electron has small (of the order of $m$)
perpendicular momentum transfer and very small (of the order of
$m^3/s$) longitudinal one. Just such kind of the events contribute
to the nondecreasing cross-section. The contribution of the rest
diagrams, describing the direct capture of the
 photon by the initial electron and exchange effects due to
the identity of the final electrons, decreases at least as
$m/\omega.$

\vspace{0.4cm}

\begin{minipage}{140 mm}
\begin{center}
\includegraphics[width=0.6\textwidth]{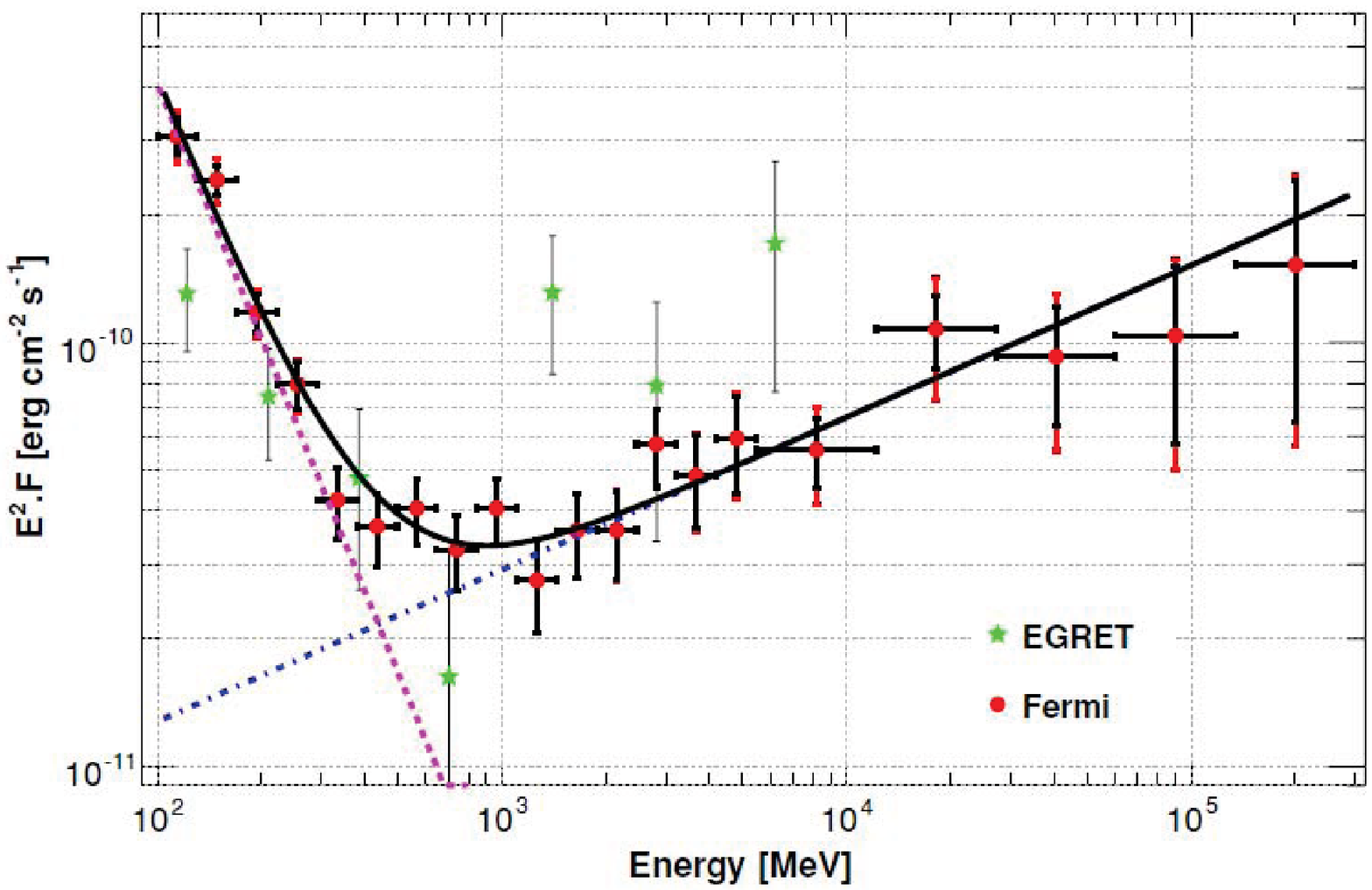}\\
\end{center}
\emph{\textbf{Fig.2}.}{ \emph{Spectral energy distribution of the
Crab Nebula renormalized to the total phase interval. The
synchrotron radiation (region up to $10^3\,MeV$) and due to
inverse Compton scattering (region upper $10^3\,MeV$) are
presented. To clarify the experimental points, statistical and
systematic errors, solid and dotted curves see Refs. \cite {AA10, K01}.}}\\
\end{minipage}

The very high-energy photon component can be contained in the
cosmic rays. For example, in Fig.$\,$2 we show the spectral energy
distribution of the Crab Nebula \cite{AA10}. In this figure the
high-energy photons in the region of $10^2-10^3 MeV$ are
synchrotron ones and in the region $10^3-10^5 MeV$ they arise due
to possibility of the inverse Compton scattering on an
ultra-energetic electrons. The analysis of their polarization is
very important to understand the remarkable features of the
cosmologically distant gamma ray bursts.

There are a few possibilities to measure the photon circular
polarization in the process (1). It is possible: i)$\,$to use
longitudinally polarized electrons and measure the asymmetry of
the cross-section at two opposite directions of the polarization,
ii)$\,$to measure the polarization of the recoil electrons,
iii)$\,$to measure the polarization of the created electrons or
positrons. The double-spin correlation effects in first two cases
decrease with the growth of the photon energy and, therefore, are
not effective for the measurement of the photon circular
polarization at the high energies. So, in this paper we
concentrate on the third experimental setup which can be realized
in the scattering of the photons on unpolarized atomic electrons
or the electron beam. Our study in some aspects is close to the
approach developed in Ref. \cite{BKGP02}, where the process (1)
with circularly polarized photons has been suggested to create
high energy polarized electrons, and in the last section we
discuss the corresponding similarities and differences in more
details.

\section{Four-rank compton tensor}

In the approximation considered here, the matrix element squared
of the process (1) is defined by a contraction of two second-rank
Lorentz tensors $V_{\mu\nu}$ and $B_{\mu\nu},$ and the
differential cross-section of this process can be written in the
form
\begin{equation}\label{3}
d\sigma=\frac{e^6}{2(2\pi)^5sq^4}V_{\mu\nu}B_{\mu\nu}d\Phi\,,
\end{equation}
$$d\Phi=\frac{d^3k_1}{2E_1}\frac{d^3k_2}{2E_2}
\frac{d^3p_1}{2\varepsilon_1}\delta(k+p-p_1-k_1-k_2)\ ,$$ where
$q=k-k_1-k_2=p_1-p,$ $E_1\,(E_2)$ is the energy of the created
electron (positron) and $\varepsilon_1$ is the energy of the
recoil electron with 4-momentum $p_1.$ The tensor $B_{\mu\nu}$ is
defined by the electron current $j_{\mu}$
\begin{equation}\label{4}
B_{\mu\nu}=j_{\mu}j^*_{\nu}\,, \ \ j_{\mu}=\bar
u(p_1)\gamma_{\mu}u(p)\,,
\end{equation}
and in the case of polarized initial electron
$$B_{\mu\nu}=\frac{1}{2}Tr(\hat p_1+m)\gamma_{\mu}(\hat p+m)(1-\gamma_5\hat W)\gamma_{\nu}\,,$$
where $W_{\mu}$  is its polarization 4-vector. Taking the trace
over the spinor indices we have
\begin{equation}\label{5}
B_{\mu\nu}=q^2g_{\mu\nu}+2(pp_1)_{\mu\nu}-2im(\mu\nu qW)\,,
\end{equation}
where the following notation is used
$$(ab)_{\mu\nu} =a_{\mu}b_{\nu}+a_{\nu}b_{\mu}\,, \ \
(\mu\nu qW) =\epsilon_{\mu\nu\lambda\rho}q_{\lambda}W_{\rho}\,,\ \
\epsilon_{1230}=1\,.$$ If the initial electron is unpolarized and
one want to measure the recoil electron polarization, it needs to
substitute
$$p\leftrightarrows p_1, \ \mu\leftrightarrows \nu, \ W\rightarrow W_1$$
in the right hand side of Eq.$\,$(5) which results simply to
change $W\rightarrow W_1,$ where $W_1$ is the polarization
4-vector of the recoil electron.

For events with arbitrarily polarized photon beam, the tensor
$V_{\mu\nu}$ in Eq.$\,$(3) can be written in terms of its Stock's
parameters $\xi_i\, \ (i=1\,,2\,,3)$ and the four-rank Compton
tensor $T_{\mu\nu\lambda\rho}$ (such that its contractions with
$q_{\mu}\,, q_{\nu}$ and $k_{\lambda}\,, k_{\rho}$ equal to zero
and which will be defined below) as follows
\begin{eqnarray}\label{6}
V_{\mu\nu}=\frac{1}{2}\big(
[e_{1\lambda}e_{1\rho}+e_{2\lambda}e_{2\rho}]
+\xi_3[e_{1\lambda}e_{1\rho}-e_{2\lambda}e_{2\rho}]+ \\ \nonumber
\xi_1[e_{1\lambda}e_{2\rho}+e_{2\lambda}e_{1\rho}]-
i\xi_2[e_{1\lambda}e_{2\rho}-e_{2\lambda}e_{1\rho}]\big)T_{\mu\nu\lambda\rho}\,,
\end{eqnarray}
where the mutually orthogonal space-like 4-vectors $e_1$ and
$e_2,$ relative to which the photon polarization properties are
defined, have to satisfy the following relations
$$e_1^2=e_2^2=-1\,, \ (e_1k)=(e_2k)=(e_1e_2)=0\,.$$

The first term inside the parentheses in r.h.s. of Eq.$\,$(6) is
in charge of the events with unpolarized photon, the second and
third ones are responsible for the events with linear photon
polarization and the last one -- for the events with the circular
polarization. The parameters $\xi_1$ and $\xi_3,$ which define the
linear polarization degree of the photon, depend on the choice of
the 4-vectors $e_1$ and $e_2,$ whereas parameter $\xi_2$ does not
depend. Because we want to investigate the events with circular
photon polarization, we can choose these 4-vectors by the most
convenient way, namely
\begin{equation}\label{7}
e_{1\lambda}=\frac{\chi_1 k_{2\lambda}-\chi_2 k_{1\lambda}}{N}\,,
\ \ e_{2\lambda}=\frac{(\lambda\,k\,k_1\,k_2)}{N}
\end{equation}
with the short notation
$$N=2\chi\chi_1\chi_2 -m^2(\chi_1^2+\chi_2^2)\,, \ \chi_{1,2}=(kk_{1,2})\,, \ \chi=(k_1k_2)\,.$$
The 4-vector $e_1$ appears kindly in the expression for the
four-rank Compton tensor $T_{\mu\nu\lambda\rho}\,,$ see Eq.(8)
below.

The polarization properties of a real photon are defined by two
orthogonal 3-vectors ${\bf n_1}$ and ${\bf n_2}.$ Each of these
two vectors are orthogonal also to 3-vector of the photon momentum
${\bf k}$, and the 4-vectors $e_1$ and $e_2$ are their covariant
generalizations. It follows from the definition of $e_1$ and $e_2$
that they have both time and space components. Adding to them
4-vector $k$ with appropriate factors (it is, in fact, the gauge
transformation), one can eliminate the time and longitudinal
(along the vector ${\bf k}$) components. Thus, in arbitrary
Lorentz system with Z-axis directed along the vector ${\bf k}$ and
3-momentum lying in the (ZX) plane, where
$$k=(\omega,0,0,\omega)\,, \ k_1=(E_1,k_{1x},0,k_{1z})\,, \
k_2=(E_2,k_{2x},k_{2y},k_{2z})\,,$$ the corresponding
transformation has the form
$$(0,{\bf n_1})=e_{1\lambda}-\frac{E_1k_{2z}-E_2k_{1z}}{N}\,k_{\lambda}\,, \
(0,{\bf n_2})=e_{1\lambda}-\frac{k_{1x}k_{2y}}{N}\,k_{\lambda},$$
where
$${\bf n_1}=(n_x\,,n_y\,,0)\,, \ {\bf n_2}=(n_y\,,-n_x\,,0)\,,$$
$$n_x=\frac{\omega[(E_1-k_{1z})k_{2x}-(E_2-k_{2z})k_{1x}]}{N}\,, \ n_y=\frac{\omega(E_1-k_{1z})k_{2y}}{N}\,,$$
$$N^2=\omega^2\big\{[(E_1-k_{1z})k_{2x}-(E_2-k_{2z})k_{1x}]^2+(E_1-k_{1z})^2k_{2y}^2\big\}\,.$$
At such transformation no observables are changed due to the gauge
invariance, that manifests itself by means of the above mention
restrictions on the tensor $T_{\mu\nu\lambda\rho}$
$$k_{\lambda}T_{\mu\nu\lambda\rho}=k_{\rho}T_{\mu\nu\lambda\rho}=0\,.$$

That is why the description of all polarization phenomena in
process (1) by means of the 4-vectors  $e_1$ and $e_2$ is
completely equivalent to the description in terms of 3-vectors
${\bf n_1}$ and  ${\bf n_2}.$ The evident advantage of the
covariant description is independence from the Lorentz system.

For events in which the created electron polarization states in
the process (1) must be determined, the Borselino diagrams lead to
following expression for the tensor $T_{\mu\nu\lambda\rho}$
\begin{eqnarray}\label{8}
T_{\mu\nu\lambda\rho}=\frac{1}{2}Tr\big\{
(\hat{k_1}+m)(1-\gamma_5\hat{S})\hat{Q}_{\lambda\mu}(\hat{k_2}-m)\hat{Q}_{\nu\rho}\big\}\,,
\\ \nonumber
\hat{Q}_{\lambda\mu}=\frac{N}{\chi_1\chi_2}e_{1\lambda}\gamma_{\mu}+\frac{\gamma_{\lambda}\hat{k}\gamma_{\mu}}{2\chi_1}
-\frac{\gamma_{\mu}\hat{k}\gamma_{\lambda}}{2\chi_2}\,,
\end{eqnarray}
where $S$ is the electron spin 4-vector, with properties
$S^2=-1\,, \ (Sk_1)=0\,.$

Let us divide $T_{\mu\nu\lambda\rho}$ into two parts: the firs
part depends on 4-vector $S$ and the second one does not
$$T_{\mu\nu\lambda\rho}=T^{(0)}_{\mu\nu\lambda\rho}+T^{(S)}_{\mu\nu\lambda\rho}\,.$$
Then we can write
\begin{eqnarray}\label{9}
T^{(0)}_{\mu\nu\lambda\rho}=T_{(\mu\nu)(\lambda\rho)} +
T_{[\mu\nu][\lambda\rho]}\,, \\ \nonumber
T^{(S)}_{\mu\nu\lambda\rho}=im\big[T_{(\mu\nu)[\lambda\rho]} +
T_{[\mu\nu](\lambda\rho)}\big]\,,
\end{eqnarray}
where we use the index notation $(\alpha\beta)$ ($[\alpha\beta]$)
to emphasize the symmetry (antisymmetry) under permutation of
indices $\alpha$ and $\beta.$ These symmetry properties (9) and
form (5) for the tensor $B_{\mu\nu}$ allow to discuss some
features of the process (1) with high energy polarized photon on
the quality level.

As we noted in the Introduction, the cross section of the process
(1) (when all particles are unpolarized) does not decrease with
the growth of the photon energy. Such behavior is caused by the
terms proportional to $s^2$ in the contraction
$T_{\mu\nu\lambda\rho}B_{\mu\nu}$ which enters differential cross
section (3). On the other hand, only symmetrical component
$2(pp_1)_{\mu\nu}$ in Eq.$\,$(5) can ensure appearance of these
terms. This simple observation suggest us that the non decreasing
spin correlations in the differential cross section in considered
case are connected only with symmetrical, relative
$(\mu\leftrightarrows\nu)$ permutation, tensors
$T_{(\mu\nu)(\lambda\rho)}$ and $T_{(\mu\nu)[\lambda\rho]}.$ The
first one describes single-spin correlations which depend on
Stock's parameters $\xi_1$ and $\xi_3$ caused by the photon linear
polarization \cite{BP71}. The second one can contribute on
condition that the polarization of the created electron (or
positron) is measured, or in other words, it describes double-spin
correlation which depends on Stock's parameter $\xi_2$ that is the
degree of the photon circular polarization. In further we will
concentrate just on this double-spin correlation that can be used
to measure parameter $\xi_2.$

The antisymmetric, under $(\mu\leftrightarrows\nu)$ permutation,
tensors $T_{[\mu\nu][\lambda\rho]}$ and
$T_{[\mu\nu](\lambda\rho)}$ have not a large physical sense in the
considered problem because they can describe the spin correlations
in the differential cross section which decrease with the energy
growth at least as $s^{-1}.$ For the full description in the such
approximation there is not enough to consider only the Borselino
diagrams and one have to account for all the rest six ones. But
these tensors are connected by the cross symmetry with the
corresponding tensors in annihilation channel which are suitable
for description of the subprocess $e^+
+e^-\rightarrow\gamma+\gamma^*,$ which is important in different
radiative return measurements \cite{RR} and where there are no
contribution of any other diagrams. That is why we give here all
the corresponding expressions in very compact form

\begin{eqnarray}\label{10}
T_{(\mu\nu)(\lambda\rho)}=\frac{2}{\chi_1\chi_2}\Big\{g_{\mu\nu}\Big[\big(\chi_1+\chi_2\big)^2g_{\lambda\rho}-
\frac{N^2}{\chi_1\chi_2}\,q^2\,e_{1\lambda}e_{1\rho}\Big]- \\
\nonumber
2\chi_1\chi_2(1+\hat{P_{\lambda\rho}})\,g_{\mu\rho}g_{\nu\lambda}
-2(k_1k_2)_{\lambda\rho}k_{\mu}k_{\nu}+ \\ \nonumber
\big(1+\hat{P_{\lambda\rho}}+\hat{P_{\mu\nu}}+\hat{P_{\lambda\rho}}\hat{P_{\mu\nu}}\big)g_{\nu\rho}\big[
k_{\mu}(\chi_2 k_{1\lambda}+\chi_1 k_{2\lambda})+N(k_{1\mu}-k_{2\mu})e_{1\lambda}\big]+ \\
\nonumber N\Big[
\frac{(k_1e_1)_{\lambda\rho}(kk_2)_{\mu\nu}}{\chi_1}-\frac{(k_2e_1)_{\lambda\rho}(kk_1)_{\mu\nu}}{\chi_1}\Big]-
\\ \nonumber
g_{\lambda\rho}\big[(\chi_1+\chi_2)(k_{12}k)_{\mu\nu}-2(m^2+\chi)k_{\mu}k_{\nu}\big]
\Big\}\,,
\end{eqnarray}
where $k_{12}=k_1+k_2$ and $\hat{P_{\alpha\beta}}$ is operator of
the $(\alpha\rightleftarrows\beta)$ permutation.

\begin{eqnarray}\label{11}
T_{[\mu\nu][\lambda\rho]}=\frac{2}{\chi_1\chi_2}\Big\{(1-\hat{P_{\mu\nu}})\Big[(\chi_1^2+\chi_2^2)
g_{\mu\lambda}g_{\nu\rho}+\frac{(\chi_1^2k_{2\mu}-\chi_2^2k_{1\mu})k_{\nu}[k_1k_2]_{\lambda\rho}}{\chi_1\chi_2}\Big]+
\\ \nonumber
(1-\hat{P_{\mu\nu}}-\hat{P_{\lambda\rho}}+\hat{P_{\mu\nu}}\hat{P_{\lambda\rho}})\Big[
\frac{N}{\chi_1\chi_2}g_{\nu\lambda}e_{1\rho}(\chi_2^2k_{1\mu}-\chi_1^2k_{2\mu})+
\\ \nonumber
\frac{\chi_1^2-(\chi_1-\chi_2)(m^2+\chi)}{\chi_1}g_{\mu\rho}k_{\nu}k_{1\lambda}+
\frac{\chi_2^2-(\chi_2-\chi_1)(m^2+\chi)}{\chi_2}g_{\mu\rho}k_{\nu}k_{2\lambda}\Big]\Big\}\,,
\end{eqnarray}
where we use notation
$[ab]_{\alpha\beta}=a_{\alpha}b_{\beta}-a_{\beta}b_{\alpha}\,.$

The spin-dependent parts of $T_{\mu\nu\lambda\rho}$ read
$$T_{[\mu\nu](\lambda\rho)}=-2(\mu\nu qS)h_{\lambda}h_{\rho}+\frac{(\mu\nu qk)}{\chi_1^2\chi_2}
\big[ (\chi_2-\chi_1)(kS)g_{\lambda\rho}-$$
\begin{equation}\label{12}
\chi_1\chi_2(Sh)_{\lambda\rho}\big]-
\frac{(kS)}{\chi_1}\big[h_{\lambda}(\mu\nu\rho
q)+h_{\rho}(\mu\nu\lambda q)\big]\,,
\end{equation}
where 4-vector $h$ is defined as
$$h=\frac{k_2}{\chi_2}-\frac{k_1}{\chi_1}\,,$$
and
$$T_{(\mu\nu)[\lambda\rho]}=\Big[\frac{(k_1k_1)_{\mu\nu}+(k_2k_2)_{\mu\nu}}{\chi_1\chi_2}-\Big(
\frac{1}{\chi_1^2}+\frac{1}{\chi_2^2}\Big)(k_1k_2)_{\mu\nu}+$$
$$ +\frac{q^2(\chi_1-\chi_2)^2+\chi_1(\chi_1^2-\chi_2^2)}{2\chi_1^2\chi_2^2}\Big](\lambda\rho kS)+$$
$$+\frac{(\lambda\rho k k_{12})}{\chi_2}\Big[(Sh)_{\mu\nu}+(Sk_2)\Big(\frac{1}{\chi_1}
-\frac{1}{\chi_2}\Big)g_{\mu\nu}\Big]+$$
$$\Big\{(\mu\lambda\rho k)\Big[\Big(\frac{(k_2S)}{\chi_2}-\frac{(kS)}{\chi_1}\Big)
\Big(\frac{k_{2\nu}}{\chi_1}-\frac{k_{1\nu}}{\chi_2}\Big)+$$
\begin{equation}\label{13}
\frac{(\chi_1-\chi_2)}{2\chi_1\chi_2^2}(q^2+2\chi_1+2\chi_2)S_{\nu}\Big]
+ (\mu\leftrightarrows\nu)\Big\}.
\end{equation}

\section{Differential cross section}

When calculating the non-decreasing (with the energy growth)
contribution to the unpolarized part of the cross section one
ought to account for the terms proportional to $s^2$  in the
contraction
$T_{\mu\nu\lambda\rho}(e_{1\lambda}e_{1\rho}+e_{2\lambda}e_{2\rho})B_{\mu\nu},$
which arise due to scalar products $(k_1p),\ (k_2p)$, and $(kp).$
To do this it is enough to use approximation
$B_{\mu\nu}=4p_{\mu}p_{\nu}$ (see Ref. [3]). Then we have
$$T_{\mu\nu\lambda\rho}(e_{1\lambda}e_{1\rho}+
e_{2\lambda}e_{2\rho})B_{\mu\nu}=
-16\Big[\frac{4m^2}{\chi_1\chi_2}(k_1p)(k_2p)+$$
\begin{equation}\label{14}
(k_1p)^2\Big(\frac{q^2}{\chi_1\chi_2}-\frac{2m^2}{\chi_2^2}\Big)
+(k_2p)^2\Big(\frac{q^2}{\chi_1\chi_2}-\frac{2m^2}{\chi_1^2}\Big)\Big]\
.
\end{equation}

The main contribution to the differential cross section, within
the chosen accuracy, gives the region of small momentum
transferred $|q^2|\sim m^2.$ In this case it is useful to
introduce the so-called Sudakov's variables \cite{G73} which are
suitable for the calculation at high energies and small momentum
transferred. These variables, in fact, define the expansion of the
final state 4-momenta on the longitudinal and transversal
components relative to the 4-momenta of the initial particles. For
the process (1) we have (see also \cite{BKGP02})
$$k_2=\alpha p^{\,'} +\beta k +k_{\perp}\,, \ q=\alpha_q p^{\,'} +\beta_q k +q_{\perp}\,,$$
$$p^{\,'}=p-\frac{m^2}{s}k\,, \ s=2(kp)\,, \ p^{\,'2}=0\,, $$
$$(k_{\perp} p)=(k_{\perp} k)=(q_{\perp}p)=(q_{\perp}k)=0\,,$$
\begin{equation}\label{15}
d^4k_2=\frac{s}{2}d\alpha d\beta d^2k_{\perp}\,, \
d^4q=\frac{s}{2}d\alpha_q d\beta_q d^2q_{\perp}\,,
\end{equation}
where the 4-vectors $k_{\perp}$ and $q_{\perp}$ are the space-like
ones, so $k_{\perp}^2=-{\bf k}^2\,, \ q_{\perp}=-{\bf q}^2,$ and
${\bf k}\,, {\bf q}$ are two-dimensional  Euclidean vectors.

The phase space of the final particles with the over-all $\delta-$
function (see Eq.$\,$(3)) can be written as
\begin{equation}\label{16}
d\Phi=\frac{s^2}{4}d\alpha d\beta d^2k_{\perp}d\alpha_q d\beta_q
d^2q_{\perp}\delta(k_2^2-m^2)\delta(k_1^2-m^2)\delta(p_1^2-m^2)\,.
\end{equation}
By using the conservation laws we derive
$$k_2^2=s\alpha\beta -{\bf k}^2\,, \ k_1^2=-s(1-\beta)(\alpha+\alpha_q)-({\bf k}+{\bf q})^2\,,$$
$$p_1^2=s\beta_q+m^2-{\bf q}^2\,, \ s\alpha=\frac{m^2+{\bf
k}^2}{\beta}\,, \ s\beta_q={\bf q}^2\,,$$
$$s\alpha_q=-\frac{m^2+{\bf k}^2}{\beta}-\frac{m^2+({\bf k}+{\bf q})^2}{1-\beta}\,,$$
and after integration over $\alpha\,, \alpha_q\,, \beta_q$ by help
of three $\delta$-functions the phase space reduces  to very
simple expression
\begin{equation}\label{17}
d\Phi= \frac{1}{4s\beta(1-\beta)}d\beta d^2{\bf k}d^2{\bf q}\,,
\end{equation}
The variable $\beta$  is the photon energy fraction that is
carried out by the positron $\beta=E_2/\omega$ (the created
electron energy $E_1=(1-\beta)\omega$). In terms of the Sudakov's
variables, the independent invariants are expressed as follows
\begin{equation}\label{18}
\chi_1=\frac{m^2+({\bf k+q})^2}{2(1-\beta)}\\, \
\chi_2=\frac{m^2+{\bf k}^2}{2\beta}\\, \ q^2=-{\bf
q}^2-\frac{m^2(m^2+{\bf k}^2)^2}{s^2\beta^2(1-\beta)^2}.
\end{equation}

Further we will consider two possible experimental situations:
i)$\,$when both the scattered (recoil) and created electron are
recorded, ii)$\,$only created electron is recorded. In the first
case we suggest that events with $|q^2|<|q^2_0|$ are not detected,
where the minimal selected momentum transfer squared $|q^2_0|$ is
of the order of $m^2.$ In the second case all events with
$|q^2|\geq|q^2_{min}|$ are included, where $|q^2_{min}|$ is the
minimal possible value of $|q^2|,$ which is defined by the second
term in the expression for $-q^2$ in Eq.$\,$(18). It is just the
longitudinal momentum transfer squared.

These two event selections give very different values for the
differential cross section. If $|q^2|$ is of the order of $m^2$
one can neglect everywhere with $q^2_{min}.$ Such procedure leads
to the cross section that depends on $q_0,$ but does not depend on
the collision energy (invariant $s$). On the other hand, when
values of $|q^2|$ for selected events begin from $q^2_{min},$ the
integration over $d^2{\bf q}$ leads to logarithmic rise of the
corresponding cross section when the collision energy increases.
This leading logarithmic contribution can be derived by the
equivalent photon method \cite{MEP} but our goal is to calculate
also the next-to-leading (constant) one.

We begin with consideration of the first experimental setup. Using
the definition of the differential cross section (Eq.$\,$(3)) and
relation (14), taking into account the phase space factor (17),
expressions for independent invariants (18) and that in considered
case
$$2(k_1p)=(1-\beta)s\,, \ 2(k_2p)=\beta s\,, \ q^2=-{\bf q}^2,$$
we obtain
\begin{equation}\label{19}
d\sigma=\frac{2\alpha^3}{\pi^2 {\bf
q}^4}\Big[2m^2\beta(1-\beta)\Big(\frac{1}{m^2+({\bf k+q})^2}-
\end{equation}
$$
\frac{1}{m^2+{\bf k}^2}\Big)^2+\frac{{\bf
q}^2[1-2\beta(1-\beta)]}{[m^2+({\bf k+q})^2][m^2+{\bf k}^2]}\
\Big]d\beta d^2{\bf k}d^2{\bf q}\ .$$

Our goal is to derive distribution on the electron (positron)
energy $\beta$ and the perpendicular momentum transfer squared
(${\bf q}^2).$ Thus, we have to integrate the r.h.s. of
Eq.$\,$(19) over $d^2{\bf k},$ and the effective values of $|{\bf
k}|$ of the order of $m.$ Because integral rapidly converges we
can take $0$ and $\infty$ as the limits of integration over $|{\bf
k}|.$ After corresponding integration the differential cross
section reads
$$ \frac{d\sigma^l}{d\beta d{\bf q}^2}=\frac{4\alpha^3}{{\bf
q}^4}\Big\{\big[1-2\beta(1-\beta)\big]\Psi_1
+2\beta(1-\beta)\Psi_2
\Big\}\ ,
$$
\begin{equation}\label{20}
\Psi_1
=\frac{1}{x}\ln\frac{x+1}{x-1}\\, \ \ \Psi_2
=1-\frac{2m^2}{{\bf q}^2}\Psi_1
\\,
\end{equation}
$$x=\sqrt{1+\frac{4m^2}{{\bf q}^2}}\ .$$

In the limited case ${\bf q}^2/m^2\gg 1\,, \Psi_1=\ln({\bf
q}^2/m^2)\,, \Psi_2=1.$ If contrary ${\bf q}^2/m^2\ll 1,$ the
expression in the braces in r.h.s. of Eq.$\,$(20) has to be
proportional to ${\bf q}^2$ due to the gauge invariance. In this
case
$$\Psi_1=\frac{{\bf q}^2}{2m^2}\Big(1-\frac{{\bf q}^2}{6m^2}\Big)\,, \ \Psi_2=\frac{{\bf q}^2}{6m^2}.$$

Elementary integration this cross section over the positron energy
fraction $\beta$
\begin{equation}\label{21}
\frac{d\sigma^l}{d{\bf q}^2}=\frac{4\alpha^3}{3{\bf
q}^4}\Big[1+2\big(1-\frac{m^2}{{\bf q}^2}\Big)\Psi_1\Big]\,,
\end{equation}
allows to find distribution over the recoil momentum $l$ in the
rest frame of the initial electron (formula (16) in
Ref.$\,$\cite{SB59}) which is connected with ${\bf q}^2$ by
relation
$${\bf q}^2+2m^2=2m\sqrt{m^2+l^2}.$$ This is a checking test of
our calculation. But further we could not integrate over $\beta$
because we have to leave the trace from the created pair.

For the pair creation in the process (1) by the high-energy photon
on the relativistic initial electron with the energy $E\gg m$ at
back-to-back collision, the scattered (recoil) electron can be
detected, in principle, by means of the circular detector which
sums all events with $\theta_{min}<\theta<\theta_{max},$ where the
scattering electron angle $\theta=|{\bf q}|/E.$ Here we bear in
mind that the scattered electron energy $\varepsilon_1$
practically does not distinguish from the initial electron one
$E$. In such experimental setup the differential cross-section
(20) ought to be integrated over the detector aperture. The
maximum and minimum values of ${\bf q}^2$ are defined by the
angular dimensions of the detector.
$${\bf q}^2_{min}=E^2\theta^2_{min}\ , \ \ {\bf q}^2_{max}=E^2\theta^2_{max}\ .$$

For analytical integration it is convenient to introduce new
variable ${\bf q}^2/m^2=4\sinh^2{z}\,,$  so that
$$\Psi_1=2z\tanh{z}, \ \Psi_2=1-\frac{z}{\sinh{z}\cosh{z}}, \
\frac{d{\bf q}^2}{{\bf q}^4}=\frac{\cosh{z}dz}{2m^2\sinh^3{z}}\
.$$ and the integration of the Eq. (20) with respect to the
azimuth angle and new variable $z$  leads to following electron
(positron) spectrum for the unpolarized case
\begin{equation}\label{22}
\frac{d\sigma}{d\beta}=2\alpha
r_0^2\big\{A(z_0)-A(z_1)+\beta(1-\beta)\big[B(z_0)-B(z_1)\big]\big\}\\,
\end{equation}
where $z_0$ and $z_1$ are the minimal and maximal values of $z\,,
\ z=Arcsinh\big(\theta E/(2m)\big)$ and functions $A(z)$ and
$B(z)$ are defined as follows
\begin{equation}\label{23}
A(z)=2z\coth{z}-2\ln(2\sinh{z})\ ,
\end{equation}
$$B(z)=\frac{2}{3\sinh^2{z}}-2z\coth{z}-\frac{2}{3}z\coth^3{z}
+\frac{8}{3}\ln(2\sinh{z}).$$ When writing the last formulae we
fixed the integration constant in such a way to both $A(z),\ B(z)
\rightarrow 0$ if $z\rightarrow\infty.$ This choice specified by
the behavior of the cross-section (16) at large ${\bf q}^2/m^2.$

The total cross-section in such experimental setup can be derived
by elementary integration over the positron energy fraction
\begin{equation}\label{24}
\sigma=2\alpha r_0^2\big[\,C(z_0)-C(z_1)\big]\\, \
C(z)=A(z)+\frac{1}{6}B(z)\ .
\end{equation}

Note, that in the $e^+e^-$ pair production by the photon on the
stationary target with arbitrary mass $M$ the quantity ${\bf
{q}^2}$ is connected with the target mass $M$ and the energy $W$
of the recoil particle in the lab. system
$${\bf {q}^2}=2M(W-M)\,, \ \ W=\sqrt{M^2+l^2}\,,$$
where $l$ is the absolute value of the recoil momentum. It means
that in the case of the atomic electron target $l=m\sinh(2z)\,,
(M=m),$ and for the very heavy target $l=2m\sinh(z), (M>>m ).$ For
the stationary target it is possible to investigate such
experimental setup when detector records all events with $l>l_0\,,
 \ l_0\sim m.$ In this case we can formally suppose the upper limit of
integration in Eq. (16) to be equal to infinity. To write the
corresponding results it is enough eliminate $A(z_1), B(z_1)$ and
$C(z_1)$ in Eqs. (22) and (24).

On the other hand, one can study the angular distribution of the
recoil electrons. It is easy to see that in this case the angle
$\vartheta$ between the photon 3-momentum and the recoil electron
one ${\bf p_1}$ is defined by relation \cite{VK73}
$$\sin^2{\vartheta}=\frac{4m^2}{4m^2 + {\bf q^2}}.$$ It means that
large ${\bf q^2}$ correspond to small recoil angles $\vartheta$
and vice-versa. In this case $\sinh{z}=\cot{\vartheta}.$

Let us consider situation when the recoil electron is not
detected. In this case we must integrate over all possible region
of variation of the variable ${\bf q}^2,$ beginning from zero. At
very small ${\bf q}^2,$ such that
\begin{equation}\label{25}
0 <{\bf q}^2<\sigma\,, \ m^6/s^2\ll\sigma\ll m^2\,,
\end{equation}
the differential cross section could be modified by substitution
in the denominator in the r.h.s. of Eq.(19) (in accordance with
definition of the cross section (3) and relation (18) for quantity
$q^2$)
 $${\bf q}^4\rightarrow\Big({\bf q}^2+\frac{m^2s_1^2}{s^2}\Big)^2\,, \ s_1=\frac{{\bf k}^2+m^2}{\beta(1-\beta)}\,,$$
 where quantity $s_1$ is the invariant mass squared of the created electron-positron pair in the
 process (1) at ${\bf q}=0.$ Besides, in the numerator we can neglect terms of the order of ${\bf q}^n$
 if the power $n>2.$ In the region $\sigma<{\bf
 q}^2<\infty$ we can use expression (19).

 In the region
 (25) the gauge invariance requires ${\bf q}^2$ - dependence of the cross section of the following form \cite {G73}
 $$\frac{{\bf q}^2}{\big({\bf q}^2+m^2s_1^2/s^2\big)^2}\,.$$
 Thus, we can perform elementary integration over $d^2{\bf q}$ in
 this region and the azimuth angle of the two-dimensional vector ${\bf k}$ and
 derive
 \begin{equation}\label{26}
\frac{d\sigma^s}{d\beta d{\bf k}^2}=\frac{2\alpha^3}{(m^2+{\bf
k}^2)^2}\Big[1-2\beta(1-\beta)+\frac{4\beta(1-\beta)m^2{\bf
k}^2}{(m^2+{\bf k}^2)^2}\Big]\Big[\ln\Big(\frac{\sigma
s^2\beta^2(1-\beta)^2}{m^2(m^2+{\bf k}^2)^2}\Big)-1\Big].
 \end{equation}

 To obtain the electron (positron) spectrum in the region (25) we
 have to integrate Eq.$\,$(26) over $d{\bf k}^2.$ The result reads
 \begin{equation}\label{27}
\frac{d\sigma^s}{d\beta}=2\alpha
r_0^2\Big\{\big[1-\frac{4}{3}\beta(1-\beta)\big]\Big[
\ln\Big(\frac{\sigma
s^2\beta^2(1-\beta)^2}{m^6}\Big)-1\Big]-2+\frac{26}{9}\beta(1-\beta)\Big\}\,.
 \end{equation}

The total electron (positron) spectrum contains also contribution
of the region $\sigma<{\bf
 q}^2<\infty.$ To derive it we integrate Eq.$\,$(20) over ${\bf q}^2$
 and obtain
 \begin{equation}\label{28}
\frac{d\sigma^l}{d\beta}=2\alpha
r_0^2\Big[2-\ln\frac{\sigma}{m^2}+2\beta(1-\beta)\Big(
\frac{2}{3}\ln\frac{\sigma}{m^2}-\frac{13}{9}\Big)\Big].
 \end{equation}

The electron spectrum in the case of the undetected recoil
electron is the sum of $d\sigma^s/d\beta$ and $d\sigma^l/d\beta$
which is defined by the well known expression
\begin{equation}\label{29}
\frac{d\sigma}{d\beta}=2\alpha
r_0^2\big[1-\frac{4}{3}\beta(1-\beta)\big]\Big[
\ln\Big(\frac{s^2\beta^2(1-\beta)^2}{m^4}\Big)-1\Big]\,.
\end{equation}
It describes the corresponding differential cross section for
$e^+e^-$ pair production by the high energy photon on elementary
electric charge. It is suitable also for pair production in the
non-screening Coulomb field (with substitution
$\alpha^3\rightarrow\alpha^3Z^2.$)

If the recoil electron is not detected we can also study the
double distribution of the positron over the energy $\omega\beta$
and perpendicular momentum ${\bf k}$ which are related with its
scattering angle $\theta$ : $\theta=2|{\bf k}|/(\beta\sqrt{s}).$
For this goal we have to integrate differential cross section (19)
over $d^2{\bf q}$ in the region ${\bf q}^2>\sigma$ and result add
to contribution (26) from the region ${\bf q}^2<\sigma.$ Such
integration of the expression (19) gives
\begin{equation}\label{30}
\frac{d\sigma^l}{d\beta d{\bf k}^2}=\frac{2\alpha^3}{(m^2+{\bf
k}^2)^2}\Big\{2\beta(1-\beta)\Big[1-\frac{6m^2{\bf k}^2}{(m^2+{\bf
k}^2)^2}\Big]-
\end{equation}
$$\Big[1-2\beta(1-\beta)+\frac{4\beta(1-\beta)m^2{\bf
k}^2}{(m^2+{\bf k}^2)^2}\Big]\ln\frac{\sigma m^2}{(m^2+{\bf
k}^2)^2}\Big\}\,.$$

In the total distribution the term $\ln\big(\sigma m^2/(m^2+{\bf
k}^2)^2\big)$  cancels and we obtain

\begin{equation}\label{31}
\frac{d\sigma}{d\beta d{\bf k}^2}=\frac{2\alpha^3}{(m^2+{\bf
k}^2)^2}\Big\{2\beta(1-\beta)\Big(1-\frac{6m^2{\bf k}^2}{(m^2+{\bf
k}^2)^2}\Big)+
\end{equation}
$$\Big[1-2\beta(1-\beta)+\frac{4m^2{\bf k}^2\beta(1-\beta)}{(m^2+{\bf
k}^2)^2}\Big]\Big(2\ln\frac{s\beta(1-\beta)}{m^2}-1\Big)\Big\}\,.$$

When integrating (31) with respect to $d{\bf k}^2,$ the first term
in the curly brackets vanishes and we come to the spectrum (29).
On the other hand we can also integrate (31) over $\beta$ and
obtain

\begin{equation}\label{32}
\frac{d\sigma}{d{\bf k}^2}=\frac{2\alpha^3}{3(m^2+{\bf
k}^2)^2}\Big[4\Big(1+\frac{m^2{\bf k}^2}{(m^2+{\bf
k}^2)^2}\Big)\ln\frac{s}{m^2}-\frac{29}{3}-\frac{44\,m^2{\bf
k}^2}{3(m^2+{\bf k}^2)^2}\Big]\,.
\end{equation}

\section{Polarization of the created electron}

The created (fast) electron polarization in the process (1)
depends on all kinematical variables and at high energies can be
written as follows

\begin{equation}\label{33}
P(\beta,{\bf
k,\,q})=m\xi_2\frac{T_{(\mu\nu)[\lambda\rho]}(e_{1\lambda}e_{2\rho}-
e_{1\rho}e_{2\lambda})4\,p_{\mu}p_{\nu}d\Phi/q^4}{T_{(\mu\nu)(\lambda\rho)}(e_{1\lambda}e_{1\rho}+
e_{2\lambda}e_{2\rho})4\,p_{\mu}p_{\nu}d\Phi/q^4}\,.
\end{equation}
Note that in this equation we can eliminate factor $d\Phi/q^4,$
but if our aim is, for example, to obtain the quantities
$P(\beta,{\bf q})$ or $P(\beta)$ and so on, we first have to
integrate both numerator and denominator over corresponding
variables and only then to take their ratio. It is obvious that
denominator is defined by the cross section and we have to
investigate the numerator (or the part of the cross section which
depends on circular polarization of the photon and longitudinal
polarization of the created electron) in different experimental
setups.

In terms of used invariants the numerator in Eq.$\,$(33) reads
(without the factor $d\Phi/q^4$)
$$16m\xi_2\Big\{\Big[\frac{(k_2p)}{\chi_1}-\frac{(k_1p)}{\chi_2}\Big]
\Big[\frac{\chi_1+\chi_2}{\chi_1\chi_2}\big[(k_2p)(kS)+\chi_1(pS)\big]$$
\begin{equation}\label{34}
-\frac{(kp)(k_2S)}{\chi_2}\Big]+\frac{q^2(\chi_2-\chi_1)(kp)(pS)}{2\chi_1\chi_2^2}\Big\}
.
\end{equation}
Further we use the covariant form of the electron polarization
4-vector, namely
\begin{equation}\label{35}
S=\frac{(kk_1)k_1-m^2k}{m(kk_1)}.
\end{equation}
It means that in the rest frame of the created electron $S=(0,{\bf
-n}),$ where ${\bf n}$ is the unit vector along the photon
3-momentum.

The used invariants are expressed through Sudakov's variables
$$2m(pS)=s\big[1-\beta-m^2/\chi_1\big], $$ $$ m(kS)=\chi_1, \
\ m(k_2S)=(k_1k_2)-m^2\chi_2/\chi_1, $$ and expression into the
circle braces in the r.h.s. of Eq.$\,$(34) has become very simple
$$ \frac{s^2q^2}{8\chi_2}\Big[\frac{1-2\beta}{\chi_1}-\frac{m^2}{\chi_1}\Big(\frac{1}{\chi_1}
-\frac{1}{\chi_2}\Big) \Big].$$

Now we can write the polarization dependent part of the cross
section
\begin{equation}\label{36}
\frac{d\sigma_{\xi}}{d\beta d^2{\bf k}d^2{\bf
q}}=-\frac{2\alpha^3\xi_2{\bf q}^2}{\pi^2q^4(m^2+{\bf
k}^2)[m^2+({\bf
k+q)}^2]}\Big[1-2\beta-
\end{equation}
$$2m^2\Big(\frac{1-\beta}{[m^2+({\bf
k+q)}^2]}-\frac{\beta}{(m^2+{\bf k}^2)}\Big)\Big]\,,$$ where we
have to take, as before for unpolarized case, $q^4={\bf q}^4$ in
the region ${\bf q}^2>\sigma$ and $[q^2={\bf q}^4+m^2s_1^2/s^2]^2$
in the region ${\bf q}^2<\sigma.$

For events with detected the scattered (or recoil) electron we can
integrate over ${d{\bf k}^2}$ and obtain the part of the double
differential cross-section in the following simple form
\begin{equation}\label{37}
\frac{d\sigma_{\xi}^l}{d\beta d^2{\bf
q}}=\frac{4\alpha^3\xi_2(1-2\beta)}{{\pi\bf
q}^4x^2}\big[\Psi_2-\Psi_1\big].
\end{equation}
Note firstly that this distribution is antisymmetrical with
respect to change $\beta$ by $1-\beta,$ whereas the polarization
independent part of the cross section (see Eq.$\,$(20)) is
symmetrical. Besides, at very small values of ${\bf q}^2$ the
cross section (37) does not depend on ${\bf q}^2$ due to the
factor $x^2$ in denominator whereas the cross section (20) has a
pole at ${\bf q}^2\rightarrow 0.$ The last feature implies that
the polarization dependent part of the spectrum in the region
${\bf q}^2<\sigma$ could not have terms which contain large
logarithm $\ln(s/m^2)$ that arises in the case of the pole-like
behavior at ${\bf q}^2\rightarrow 0.$

The created electron polarization along the direction ${\bf -n}$,
in its rest frame,  is defined by the relation
$$P=\xi_2\,P(\beta, {\bf q}^2)
=d\sigma_{\xi}^l/d\sigma^l,$$ so that the polarization transfer
coefficient
\begin{equation}\label{38}
P(\beta, {\bf
q}^2)=\frac{(1-2\beta)\big(\Psi_2-\Psi_1\big)}{x^2\big[(1-2\beta(1-\beta))\Psi_1
+2\beta(1-\beta)\Psi_2\big]}.
\end{equation}
The quantity $P(\beta, {\bf q}^2)$ is antisymmetric relative to
change $\beta\rightarrow 1-\beta,$ and its magnitude is of the
order unit inside a wide region of the kinematic variables. This
circumstance allows to measure even a rather small values of
circular polarizations.

If the scattered electrons are recorded by narrow circular
detector we have to integrate over the detector aperture as
described above for the unpolarized case. This procedure results
$$P(\beta)=\frac{(1-2\beta)\big[D(z_0)-D(z_1)\big]}{A(z_0)-A(z_1)+\beta(1-\beta)\big[B(z_0)-B(z_1)\big]},$$
\begin{equation}\label{39}
D(z)=2z[tanh(z)-coth(2z)].
\end{equation}
If all recoil momenta with $l>l_0$ are recorded then polarization
$P(\beta)$ can be derived with the same rules as it is described
at the end of Sec.$\,$3, namely one has to eliminate $A(z_1),
B(z_1)$ and $D(z_1)$ in Eq.$\,$(39) and use $l_0=2m\sinh(z_0).$ If
the angular distribution of the recoil electron is measured then
one has to use $\sinh z =\cot\vartheta.$

Consider now experimental setup without detection of the scattered
(or recoil) electron. Our goal is to obtain the double
distribution of the created electron polarization $P(\beta, {\bf
k}^2)$ and the spectrum-like one $P(\beta)$ by analogy with
Eqs.$\,$(38) and (39). Besides we can also investigate the
corresponding distribution over variable ${\bf k}^2.$ In these
cases we must take into account the contributions of both regions
${\bf q}^2>\sigma$ and ${\bf q}^2<\sigma.$

Integration of the r.h.s of Eq.$\,$(36) with respect to $d\,^2{\bf
q}$ over the region ${\bf q}^2>\sigma$ and the azimuth angle of
the vector ${\bf k}$ gives
\begin{equation}\label{40}
\frac{d\sigma^l_{\xi}}{d\beta d{\bf k}^2}=\frac{2\alpha^3
\xi_2({\bf k}^2-m^2)}{(m^2+{\bf k}^2)^3}\Big[\ln\Big(\frac{\sigma
m^2}{(m^2+{\bf k}^2)^2}\Big)\big(1-2\beta\big)+2(1-\beta)\Big]\,.
\end{equation}
We see that this part of the cross section has not a definite
symmetry relative the change $\beta\rightarrow(1-\beta).$ The
corresponding contribution of the region ${\bf q}^2<\sigma$ is
\begin{equation}\label{41}
\frac{d\sigma^s_{\xi}}{d\beta d{\bf k}^2}=\frac{2\alpha^3
\xi_2({\bf k}^2-m^2)(1-2\beta)}{(m^2+{\bf
k}^2)^3}\Big[1-\ln\Big(\frac{\sigma
s^2\beta^2(1-\beta)^2}{m^2(m^2+{\bf k}^2)^2}\Big)\Big]\,.
\end{equation}

In the sum of (40) and (41) the auxiliary parameter $\sigma$
cancels in the same manner as it takes place for unpolarized part
of the cross section and we have
\begin{equation}\label{42}
\frac{d\sigma_{\xi}}{d\beta d{\bf k}^2}=\frac{2\alpha^3 \xi_2({\bf
k}^2-m^2)}{(m^2+{\bf
k}^2)^3}\Big\{\Big[1-2\ln\Big(\frac{s\beta(1-\beta)}{
m^2}\Big)\Big]\big(1-2\beta\big)+2(1-\beta)\Big\}\,.
\end{equation}
Now we can write down the total distributions over $\beta$ and
over ${\bf k}^2.$ The elementary integrations give
\begin{equation}\label{43}
\frac{d\sigma_{\xi}}{d\beta}=0\,, \ \frac{d\sigma_{\xi}}{ d{\bf
k}^2} = \frac{2\alpha^3 \xi_2({\bf k}^2-m^2)}{(m^2+{\bf
k}^2)^3}\,.
\end{equation}

Having different distributions for both, polarization dependent
and polarization independent parts of the cross section we can
define the respective polarizations of the created electron
$P(\beta)\,, \ P({\bf k}^2)$ and $P(\beta, {\bf k}^2)$ by taking
the corresponding ratios. Note firstly that $P(\beta)$ go to zero
because $d\sigma_{\xi}/d\beta=0\,.$ Polarization $P({\bf k}^2),$
that is ratio of the right hand sides of Eqs.$\,$(43) and (32)
without factor $\xi_2$, decreases logarithmically with the rise of
the photon energy because $d\sigma_{\xi}/d{\bf k}^2$ does not
contain logarithmic contribution. Polarization $P(\beta\,, {\bf
k^2})$ (the ratio of the right hand sides of Eqs.$\,$(42) and
(31)) at very high energies goes to limit which does not depend on
the energy
\begin{equation}\label{44}
P(\beta\,, {\bf k^2})|_{s\rightarrow\infty}=\frac{(m^4-{\bf
k}^4)(1-2\beta)}{(m^2+{\bf k}^2)^2-2\beta(1-\beta)(m^4+{\bf
k}^4)}\,.
\end{equation}

The quantity $P(\beta)$ vanishes (with the accuracy of $m/\omega$)
if we take into account all events with $0<{\bf k}^2<\infty.$ But
the elimination the region of the very small values of ${\bf k}^2$
increases (in absolute value) the events number which depends on
the photon circular polarization and decreases the unpolarized
events number. Therefore, the created electron polarization can be
determined with high efficiency by the spectrum distribution of
the created electron (or positron) using the following constraint
on the event selection
$${\bf k}^2>{\bf k}_0^2\,, $$
where ${\bf k}_0^2$ is of the order of a few $m^2.$ The above
restriction means that events with very small angles of the
created electron and positron are excluded.

The simple calculation gives for the electron polarization in such
experimental setup
\begin{equation}\label{45}
P(\beta, {\bf k}^2_0)=\frac{y(1+y)A(s,\beta)}{B(y,s,\beta)}\,, \
y=\frac{{\bf k}^2_0}{m^2}\,,
\end{equation}
where
$$A(s,\beta)=2(1-\beta)+\Big[1-2\ln\Big(\frac{s\beta(1-\beta)}{m^2}\Big)\Big]\big(1-2\beta\big)\,,$$
$$B(y,s,\beta)=\Big[(1+y)^2-\frac{2}{3}\beta(1-\beta)(2+3y+3y^2)\Big]2\ln\Big(\frac{s\beta(1-\beta)}{m^2}\Big)-$$
$$-(1+y)^2+\frac{4}{3}\beta(1-\beta)(1+3y^2)\,.$$

\section{numerical results and discussions}

In this section we demonstrate some numerical estimates for the
created electron polarization (along the photon 3-momentum
direction in its rest frame) for different experimental
situations. Together with polarizations we show the corresponding
unpolarized parts of the cross section for which we always use
units $[\mu b]$ or $[\mu b/MeV^2].$ The results for different
experimental setups with detection of the recoil (or scattered)
electron are shown in Figs.$\,$3--6 and without detection -- in
Figs.$\,$7,8. All curves in these figures are correct when the
condition $s>>{\bf q^2, k^2}, m^2$ is satisfied, and we suggest
that the minimal value of the recoil 3-momentum always is of the
order of $m.$ In this case different unpolarized differential
cross sections are symmetric with respect to change
$\beta\rightarrow(1-\beta)$ whereas the polarizations are
antisymmetric.

Note that all curves in Figs.$\,$3,5,6 do not depend on the
collision energy (only the above mentioned constraints on values
$s, {\bf q}^2$ and $m^2$ are suggested correct), and the curves in
Fig.$\,$4 depend. The reason is that in Fig.$\,$4 we give the
corresponding distributions for events in c.m.s. with the fixed
scattered electron angles (but not ${\bf q}^2$). Inside the used
accuracy these angles are expressed via the perpendicular momentum
transferred and the initial electron energy by simple relation
${\bf q}^2=\theta^2\,E^2, \ E=\sqrt{s}/2.$ It means that at fixed
$\theta$ the value of ${\bf q}^2$ increases as the second power of
the energy but, as it follows from Fig.$\,$3, the cross section
decreases very quickly with the growth of ${\bf q}^2.$

\vspace{0.5cm}

\begin{minipage}{150 mm}
\begin{center}
\includegraphics[width=0.45\textwidth]{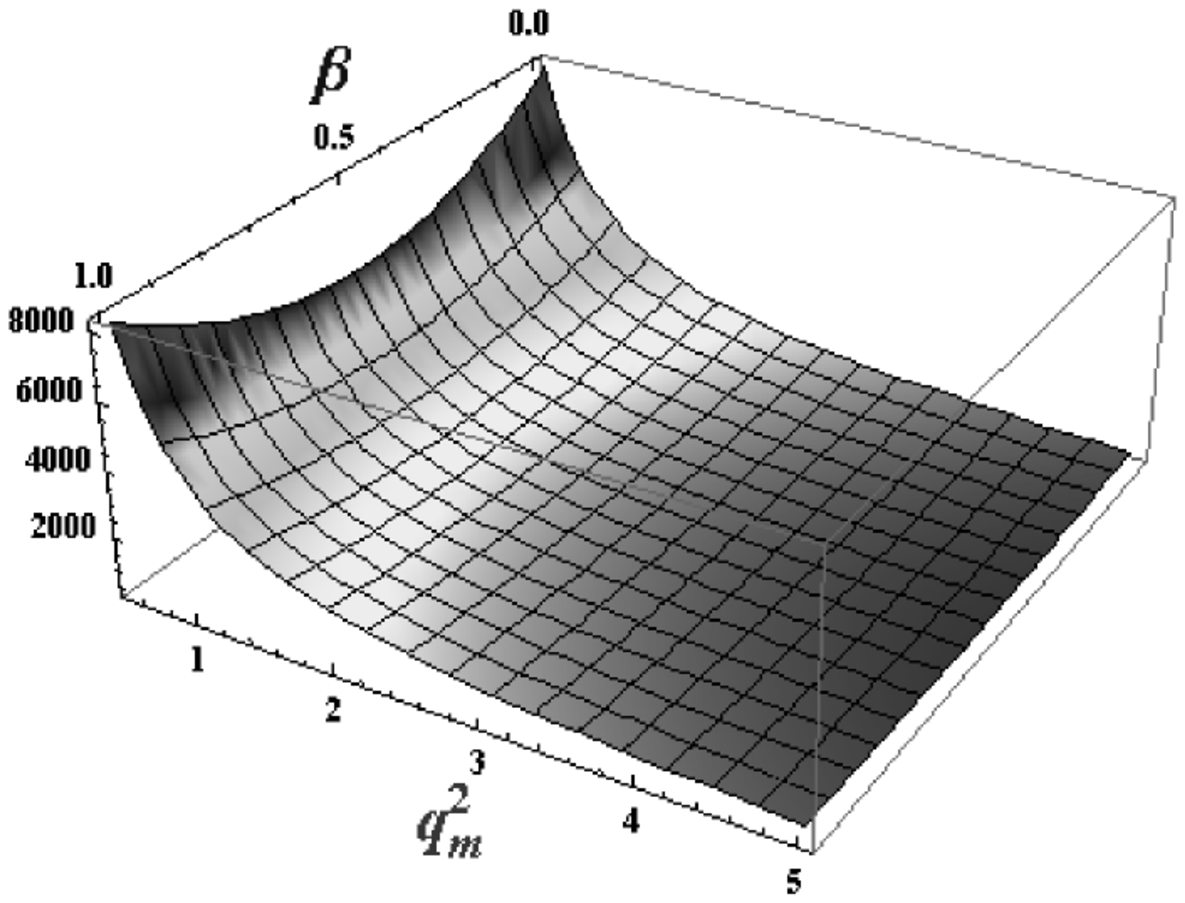}
 \hspace{0.4cm}
\includegraphics[width=0.45\textwidth]{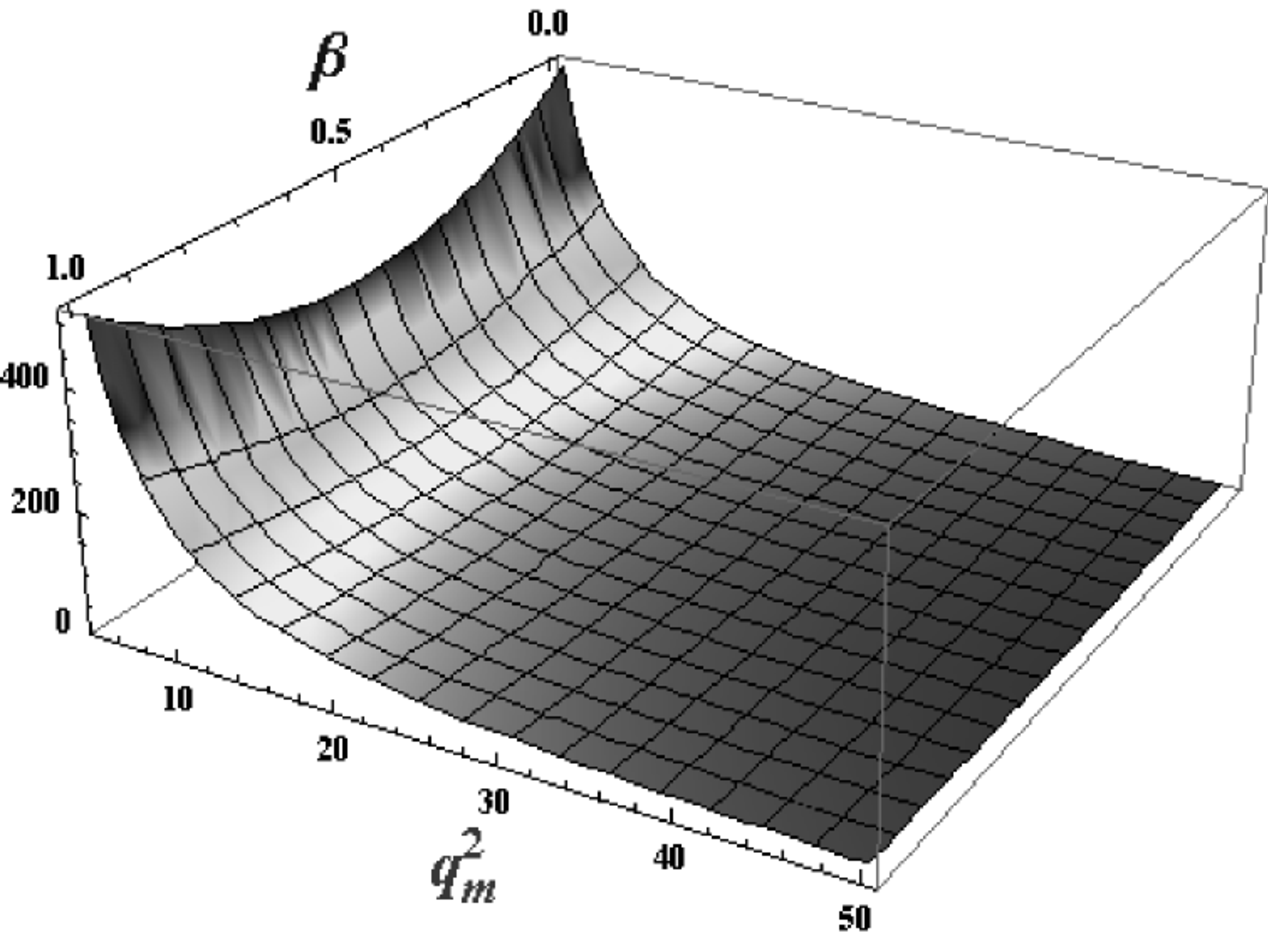}
\vspace{0.4cm}
\includegraphics[width=0.45\textwidth]{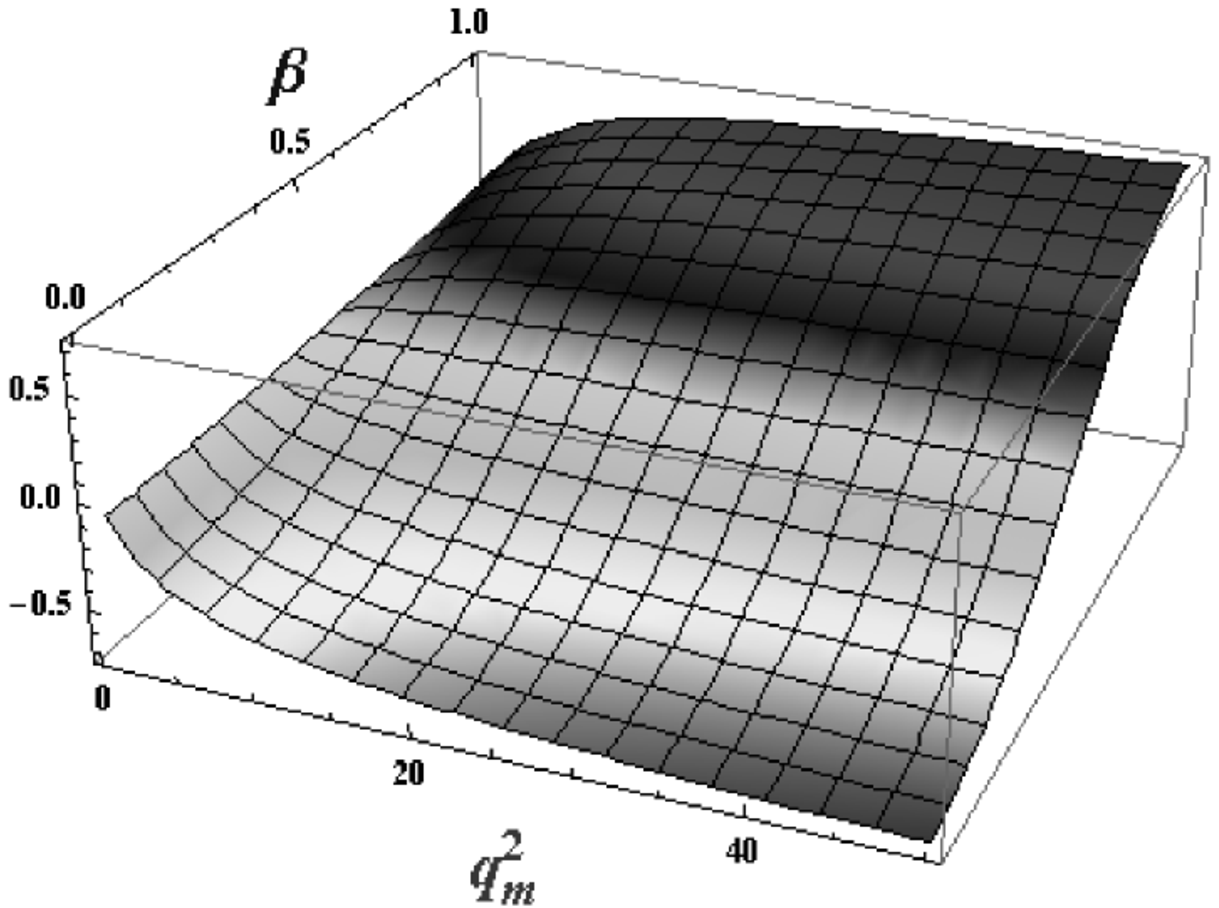} \\
\end{center}
\emph{\textbf{Fig.3.}} {\emph{Double differential cross section
(two upper graphs) as defined by Eq.(20) and the respective
distribution for the electron polarization given by Eq.(38). The
energy fraction of the electron is
$\omega(1-\beta)$ and $\,\,q^2_m={\bf q}^2/m^2.$}}\\
\end{minipage}

To imagine the energy dependence of the curves in Fig.$\,$4 we
perform corresponding calculations also at $E=1\,GeV$ and
$10\,GeV$ and can to say that polarization practically do not
depend on the energy whereas the cross section lost more then
three orders when the energy goes from $100 MeV$ to $10\, GeV.$

\vspace{0.5cm}

\begin{minipage}{150 mm}
\begin{center}
\includegraphics[width=0.4\textwidth]{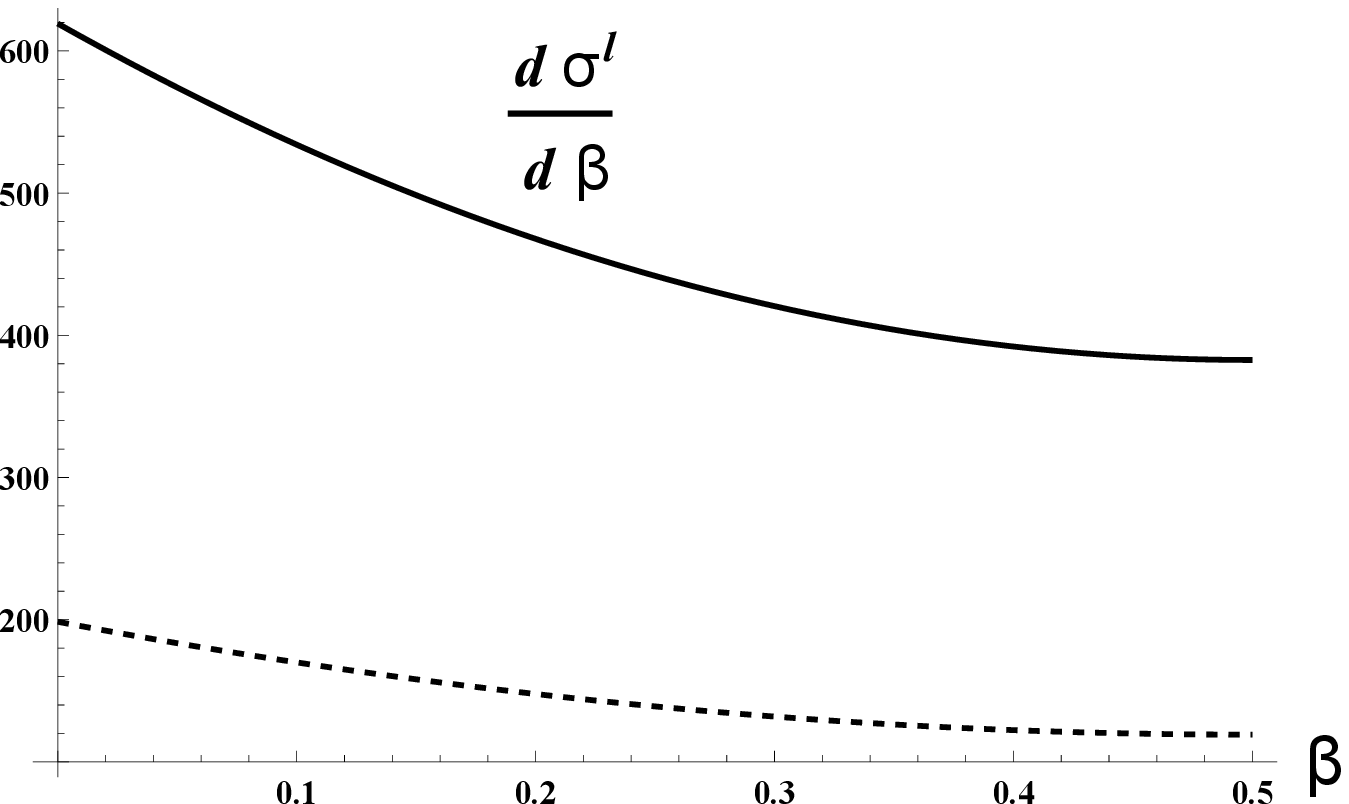}
 \hspace{0.4cm}
\includegraphics[width=0.4\textwidth]{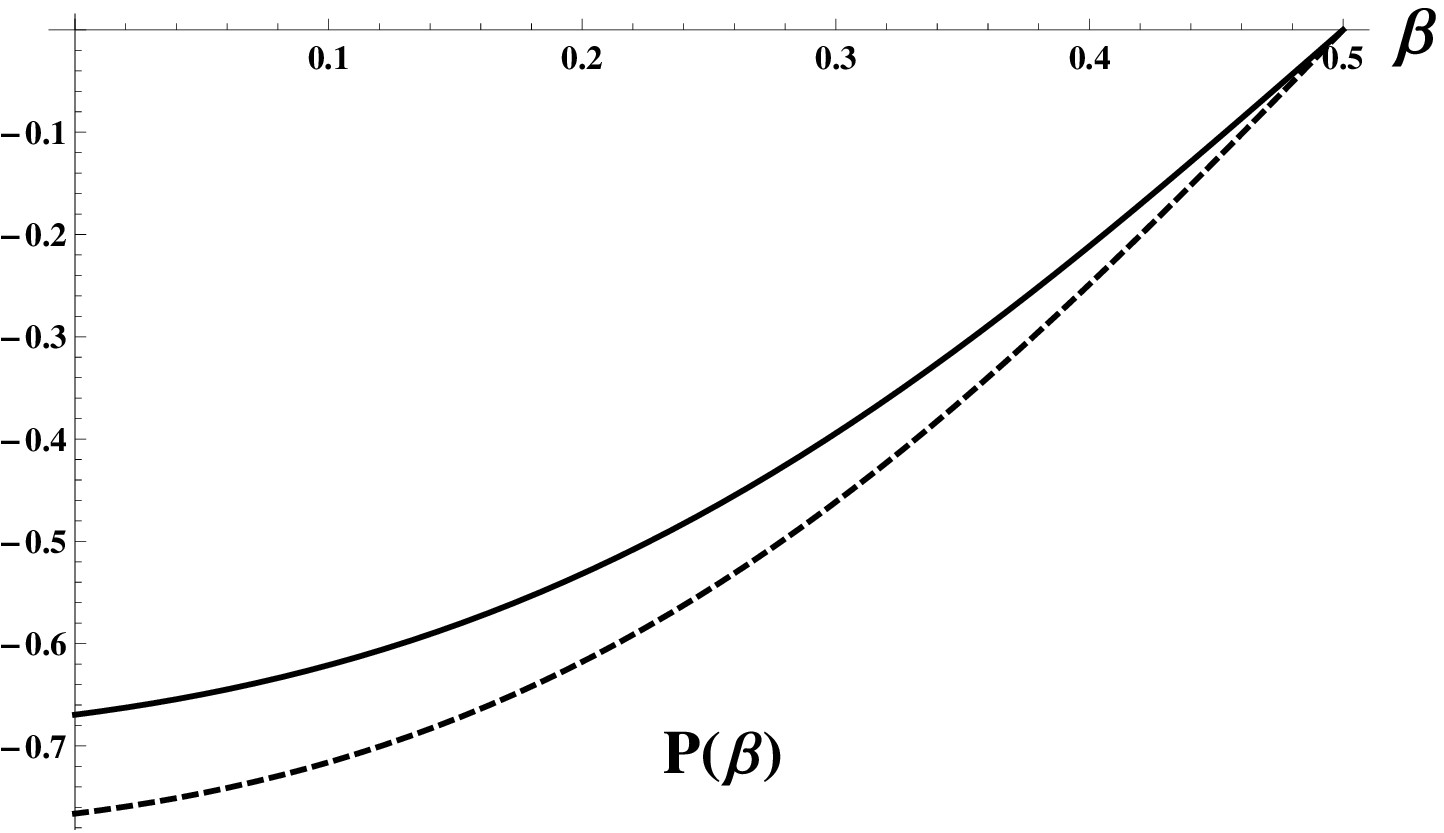}\\
\end{center}
\emph{\textbf{Fig.4.}} {\emph{The unpolarized part of the cross
section and polarization of the created electron given by
Eqs.$\,$(22) and (39) in the reaction c.m.s. (with
$z_0=Arcsinh(\theta_{min}E/2m)$ and
$z_1=Arcsinh(\theta_{max}E/2m)$) at $E=100\,MeV$ for events with
minimal electron scattering angles $\theta_{min}=1^o$ (solid
curves), $\theta_{min}=2^o$ (dotted curves) and
$\,\theta_{max}=6^o$ in both cases.
 }}\\
\end{minipage}

There is absolutely different picture of the angular distribution
of the recoil electrons for events in the rest frame. In this case
the relation between the scattered angle and the perpendicular
momentum transferred does not contain collision energy. Therefore
the curves in Fig.5 are the same at above mentioned energies.
Because in the rest frame $${\bf q}^2=4m^2cot^2(\vartheta)$$ the
cross section decreases with the growth of the recoil electron
scattering angle $\theta.$

\vspace{0.5cm}

\begin{minipage}{140 mm}
\begin{center}
\includegraphics[width=0.4\textwidth]{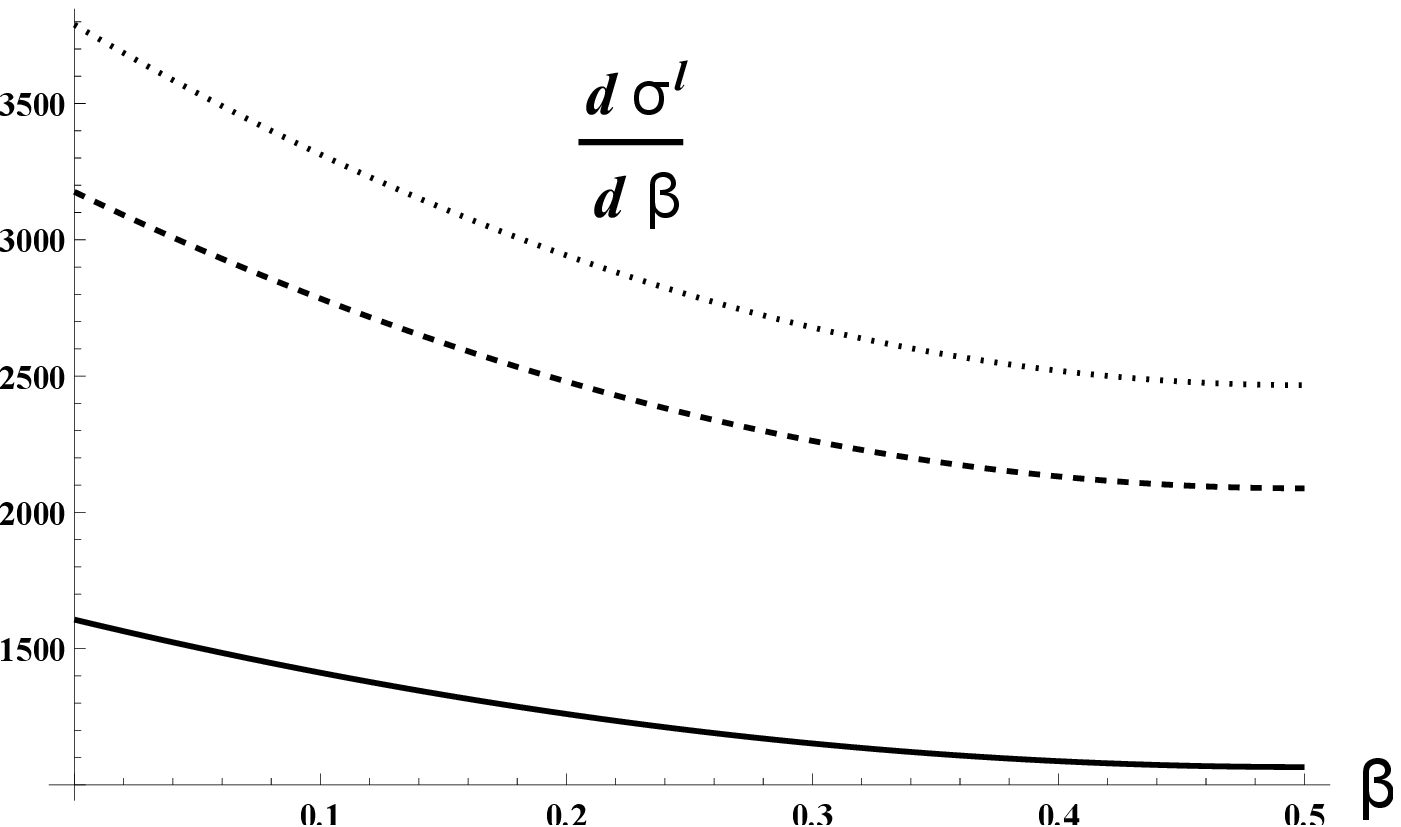}
 \hspace{0.4cm}
\includegraphics[width=0.4\textwidth]{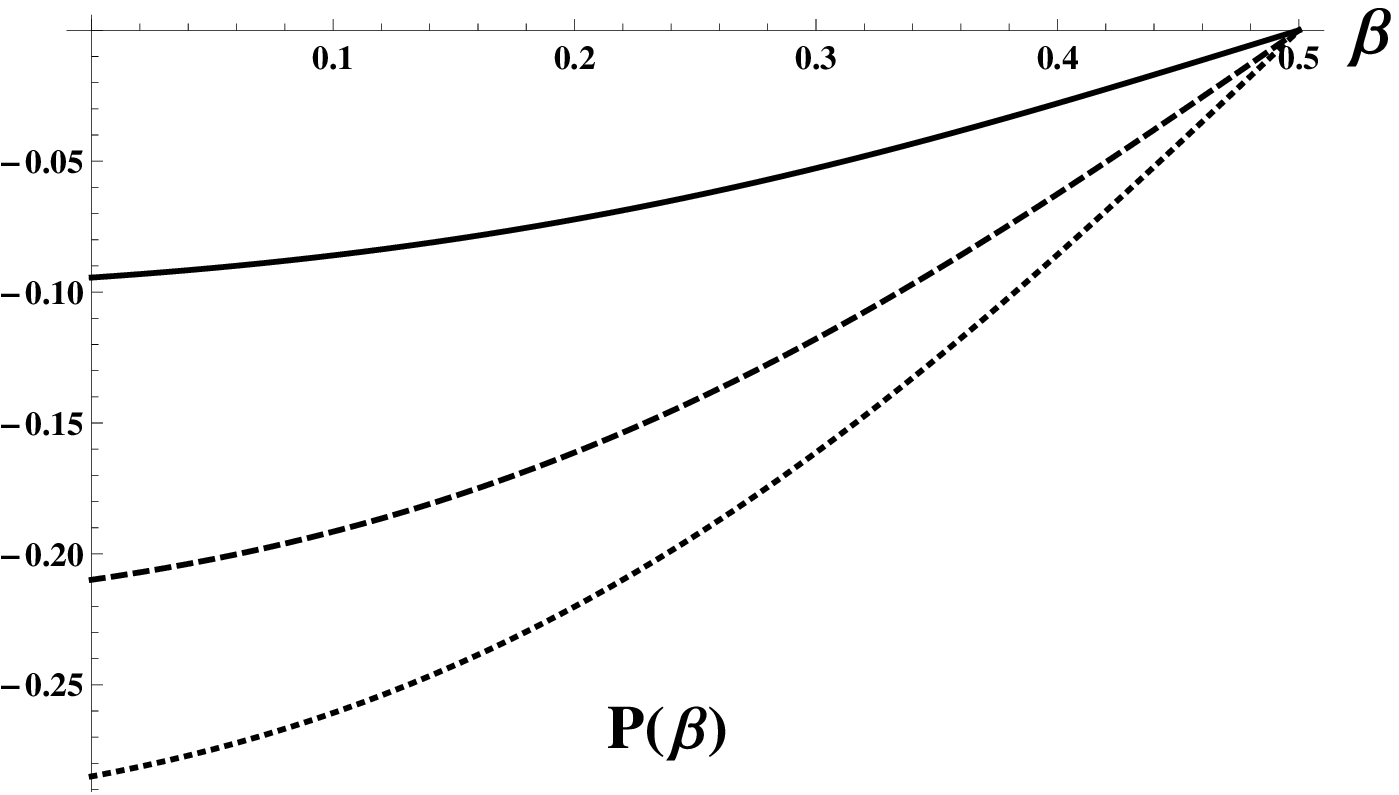}\\
\end{center}
\emph{\textbf{Fig.5.}} {\emph{The same quantities as in Fig.$\,$4
but for events in the rest frame of the initial electron with
$z_0=Arcsinh[\cot(\vartheta_{max})]$ and
$z_1=Arcsinh[\cot(\vartheta_{min})]$ in Eqs.$\,$(22) and (39). We
use $\vartheta_{max}=75^o$ and $\vartheta_{min}=60^o\,$ (solid
lines), $\theta_{min}=30^o\,$ (dashed
lines), $\vartheta_{min}=5^o\,$ (dotted lines). }}\\
\end{minipage}

To be complete with description of events at recorded the
scattered (or recoil) electron we also give in Fig.$\,$6 the
unpolarized cross section and polarization at different values of
the minimal magnitude of the recoil electron 3-momentum.

\vspace{0.4cm}

\begin{minipage}{140 mm}
\begin{center}
\includegraphics[width=0.4\textwidth]{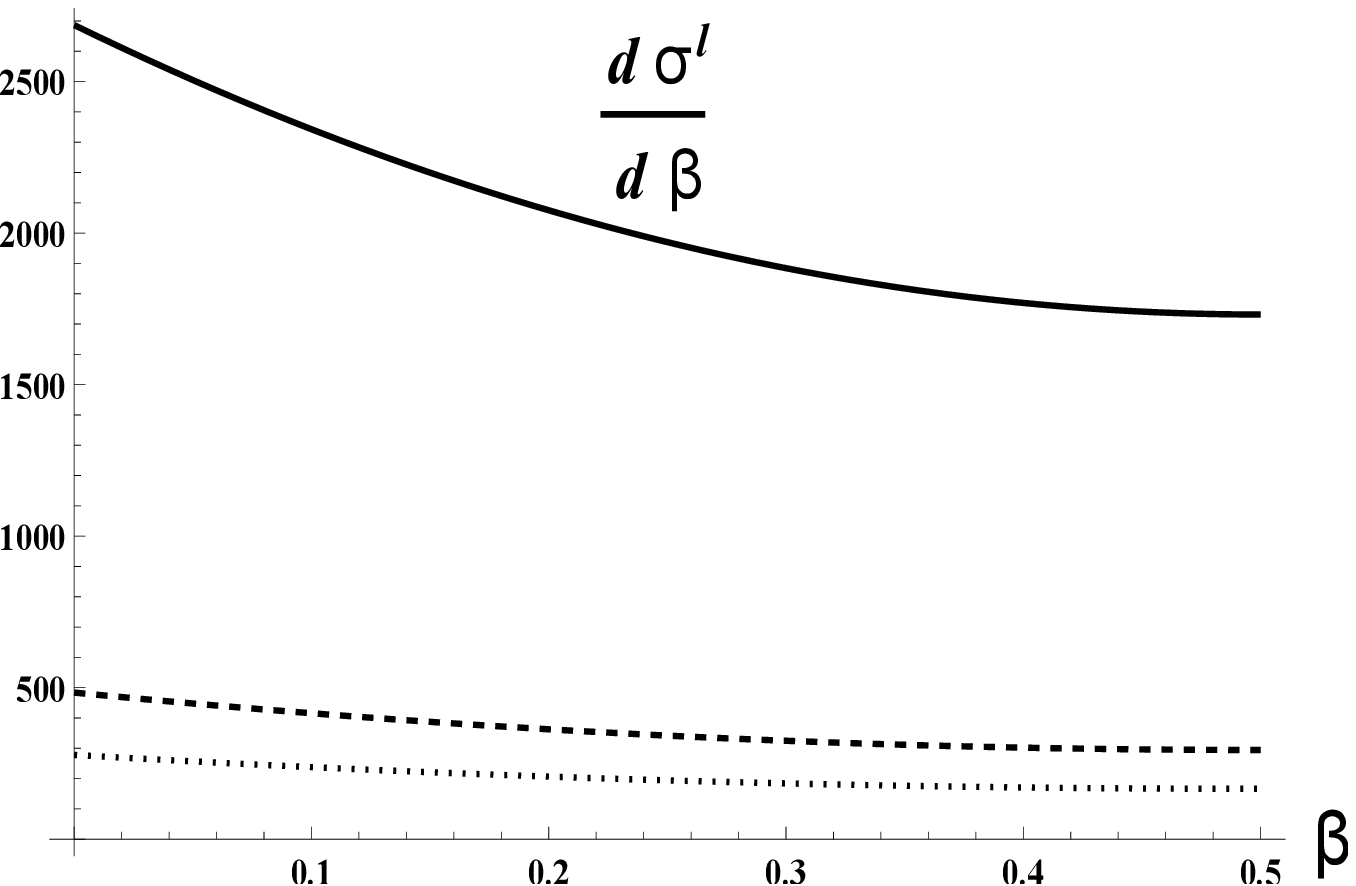}
 \hspace{0.4cm}
\includegraphics[width=0.4\textwidth]{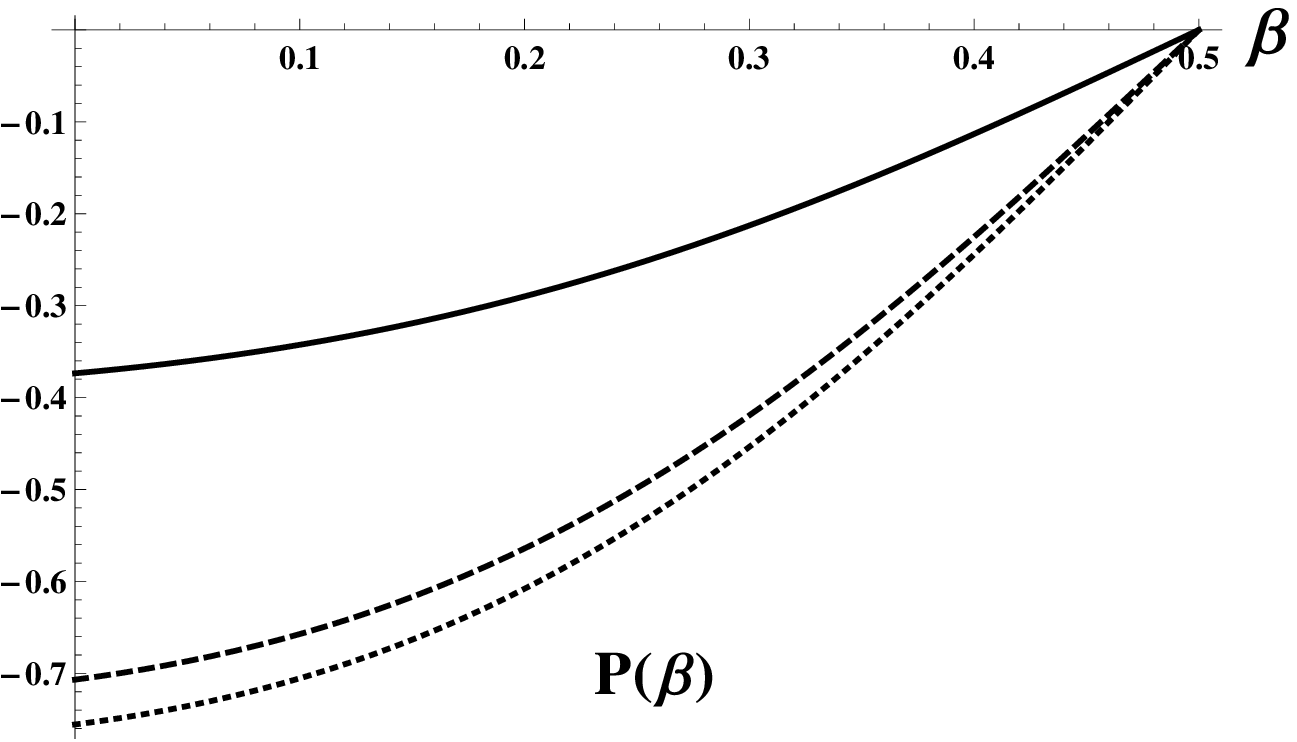}\\
\end{center}
\emph{\textbf{Fig.6.}} {\emph{The same quantities as in Fig.$\,$5
but at $z_0=Arcsinh(l_0/m)/2$ and $A(z_1)=B(z_1)=D(z_1)=0$ in
Eqs.$\,$(22) and (39);  $l_0=m$ (solid curves), $l_0=10\,m$
(dashed
curves), and $l_0=20\,m$ (dotted curves). }}\\
\end{minipage}

As concern the experimental setup without detection of the recoil
(or scattered) electron, the corresponding events include, by
definition, all values of ${\bf q}^2$ from zero ones. In this
case, the dependence of the differential cross section
$d\sigma/d\beta d^2{\bf k}d^2{\bf q}$ on the collision energy
arises due to $1/q^4$ factor in Eq.$\,$(3). After integration of
the cross section relative $d^2{\bf q}$ this dependence leaves a
trace as a term enhanced by the logarithmic factor in Eq.$\,$(31).
They say that such integrated cross section increases
logarithmically with growth of the energy.

In Fig.~8 we show the differential cross section and polarization
of the electron as a function of the positron perpendicular
momentum only. As we pointed out above, the cross section
increases logarithmically with the energy whereas the polarization
decreases. In spite of this circumstance the polarization can be
measured using such kind of distribution up to energies $ s=1
GeV^2,$ because the corresponding event number is large enough.
The more advantageous situation takes place when events with
$0<{\bf k}^2<{\bf k}^2_0$ are excluded and then the polarization
increases with the collision energy (it is demonstrated in
Fig.~9). The unpolarized cross section in this figure is not given
in text, but can be derived by integration of the cross section
(31) with respect to ${\bf k}^2$ from ${\bf k}^2_0$ up to
$\infty.$

\begin{minipage}{150 mm}
\begin{center}
\includegraphics[width=0.4\textwidth]{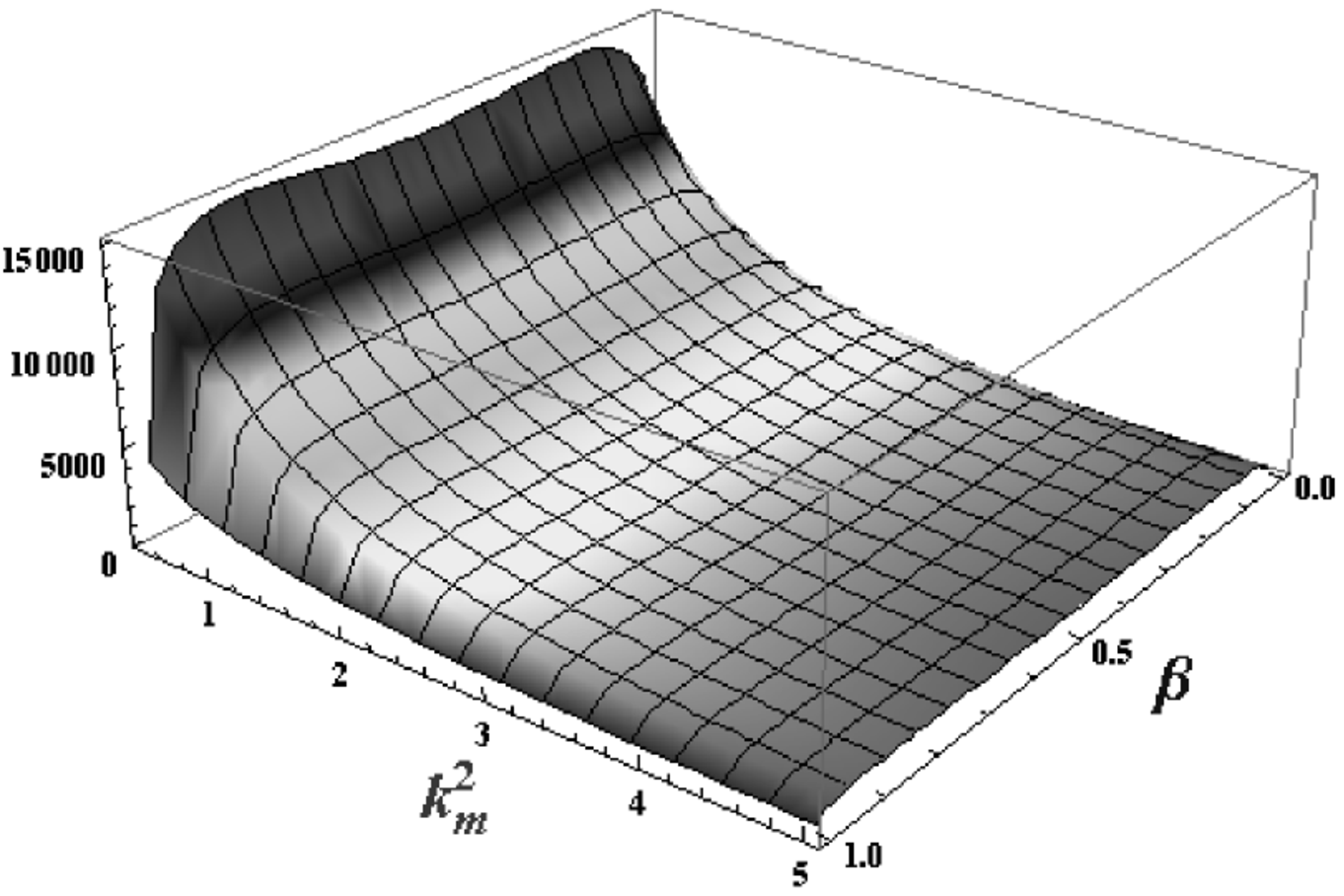}
 \hspace{0.4cm}
\includegraphics[width=0.4\textwidth]{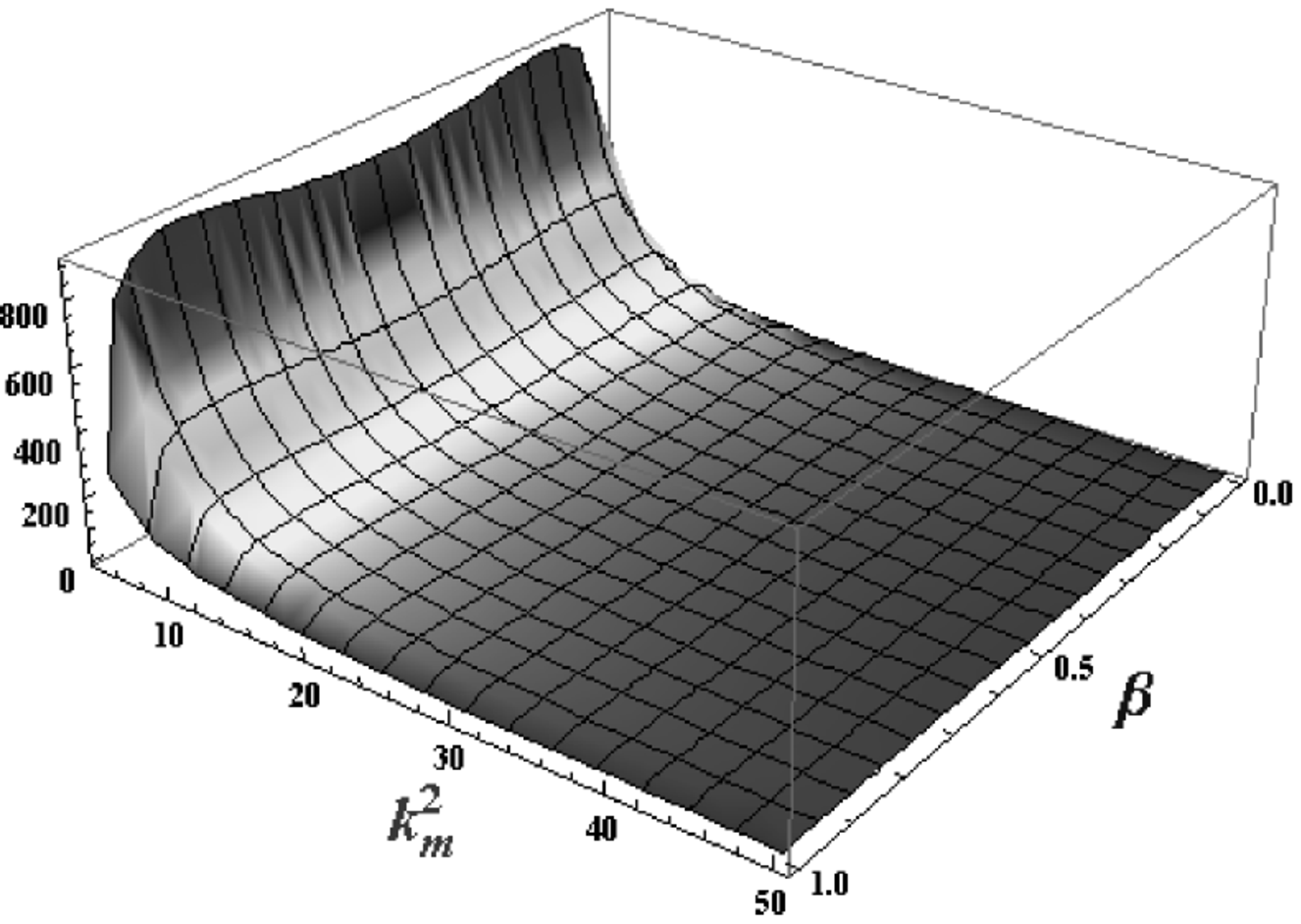}
\vspace{0.4cm}
\includegraphics[width=0.4\textwidth]{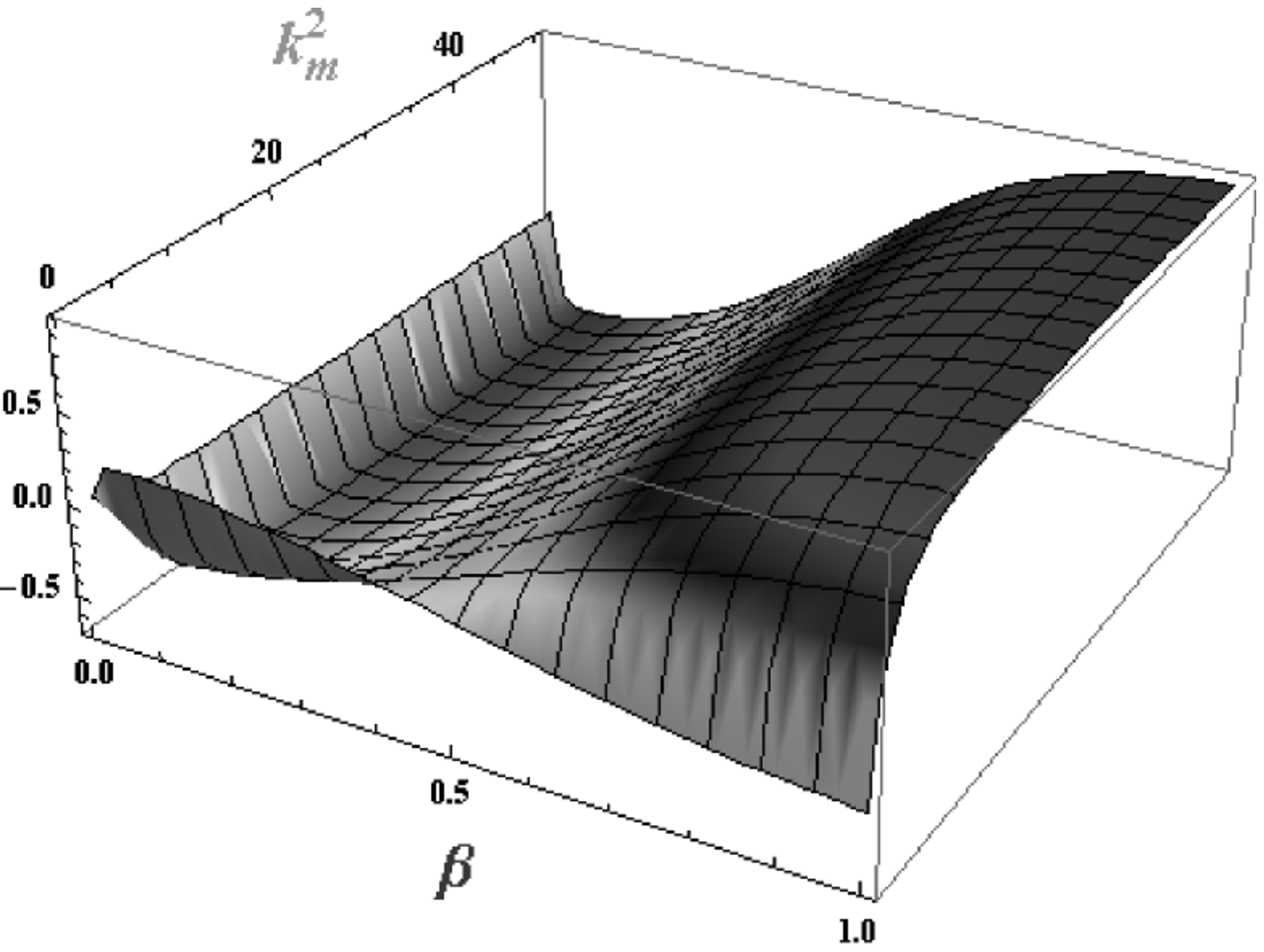} \\
\end{center}
\emph{\textbf{Fig.7.}} {\emph{Double differential cross section
(two upper figures) as defined by Eq.~(31) and the respective
distribution for the electron polarization, that is the ratio of
the right hand side of Eq.~(42) at $\xi_2=1$ to the cross section
(31), at $s=300\,MeV^2\,,$~ $k^2_m={\bf k}^2/m^2.$ }}\\
\end{minipage}

Let us compare our developed approach and obtained results with
the corresponding investigations in Ref. \cite{BKGP02}. Note first
that in both papers only the Borselino diagrams have been taken
into account for theoretical description of the process (1) and
the Sudakov's variables have been used. In Ref.~\cite{BKGP02} the
polarization of both components of the created pair is considered
but we concentrated on the polarization of the fast electron only.
In Ref.~\cite{BKGP02} the calculations were performed in the
leading logarithmic approximation using the equivalent photon
method whereas our results include also contribution which does
not depend on the energy. We consider different event selections,
particularly distributions on the recoil electron variables, which
can not be studied by the method used in Ref.~\cite{BKGP02}.

Thus, we must compare formula (14) in Ref.~\cite{BKGP02} with
coefficient at $\ln(s/m^2)$ in our unpolarized (Eq.~(26)) and
polarized (Eq.~(41))  cross sections caused by small values of
${\bf q}^2<\sigma.$ We see first that our unpolarized cross
section is twice as compared with one in Ref.~\cite{BKGP02}. It
means that we perform the spin summation. We also use  polarized
cross section that has to be twice as compared with one in
Ref.~\cite{BKGP02} (if suppose $\xi=\lambda\,,
 \ \delta_-=1$). But we see that this is not so. The reason is
 that we take different parametrization for the electron
 polarization 4-vector ( see Eq.~(35) in our work and Eq.~(12) in
 Ref.~\cite{BKGP02}). Let $\tilde S$ is the polarization 4-vector used
 in Ref.~\cite{BKGP02}. Then we have in our notation
 $$ 2m(p\tilde S)=s(1-\beta)\,, \ m(k\tilde S)=\chi_1-\frac{m^2}{1-\beta}\,, \
 m(k_2\tilde S)=(k_1k_2)-\frac{m^2\beta}{1-\beta}\,.$$

\vspace{0.5cm}

\begin{minipage}{150 mm}
\begin{center}
\includegraphics[width=0.4\textwidth]{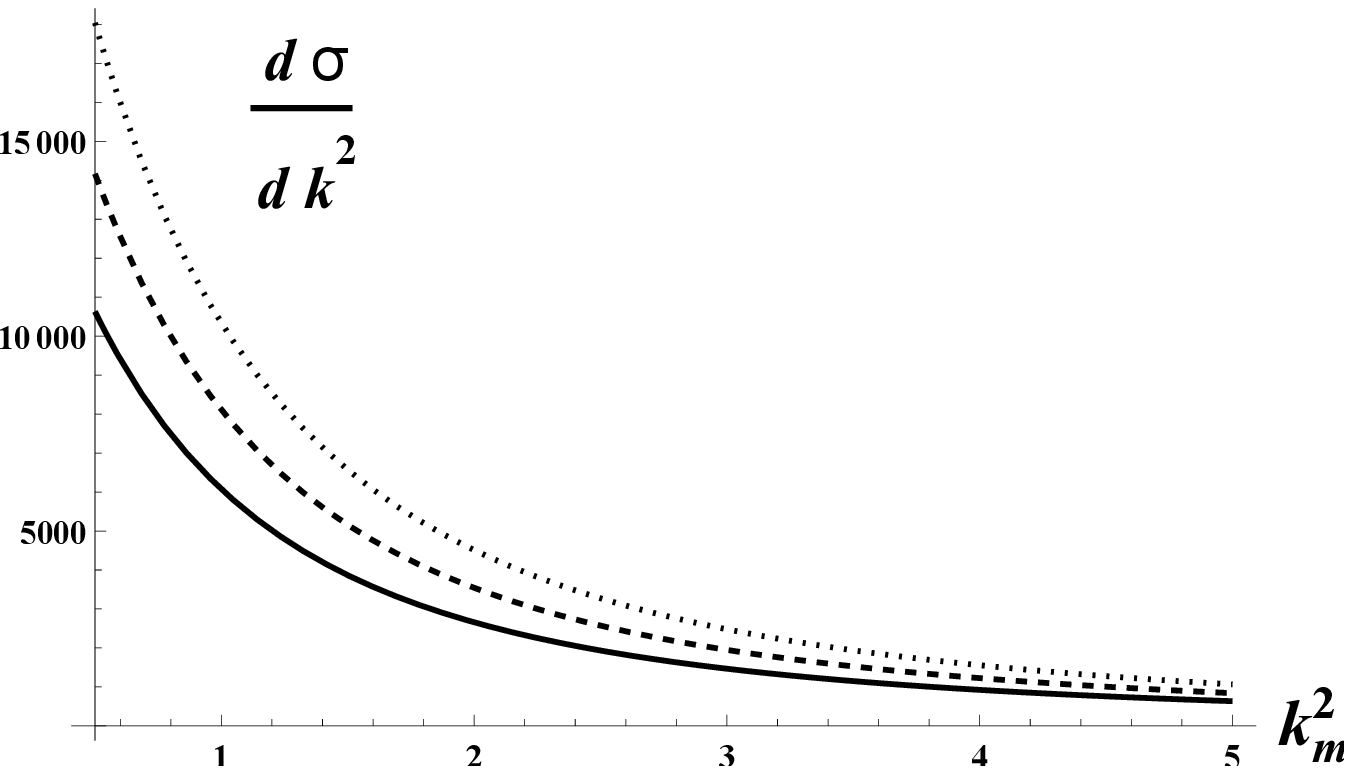}
 \hspace{0.4cm}
\includegraphics[width=0.4\textwidth]{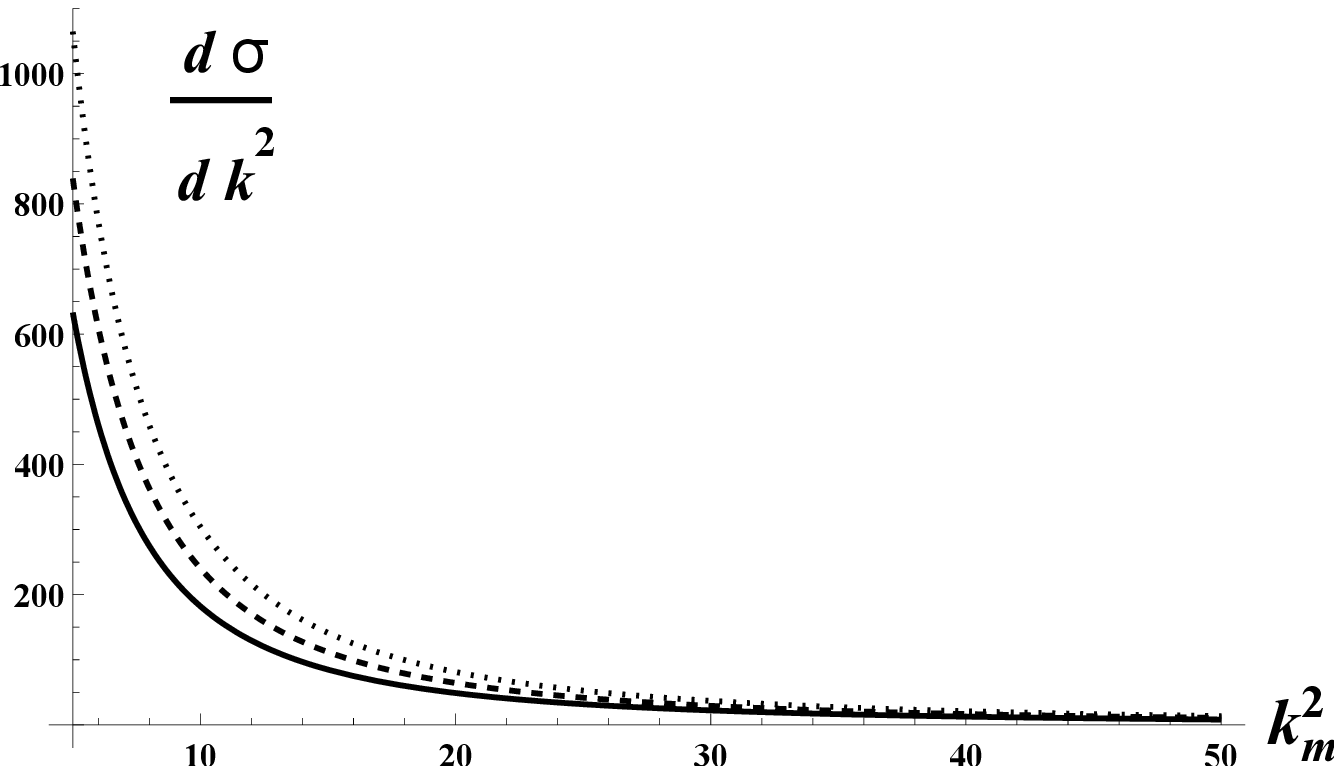}

\vspace{0.4cm}

\includegraphics[width=0.4\textwidth]{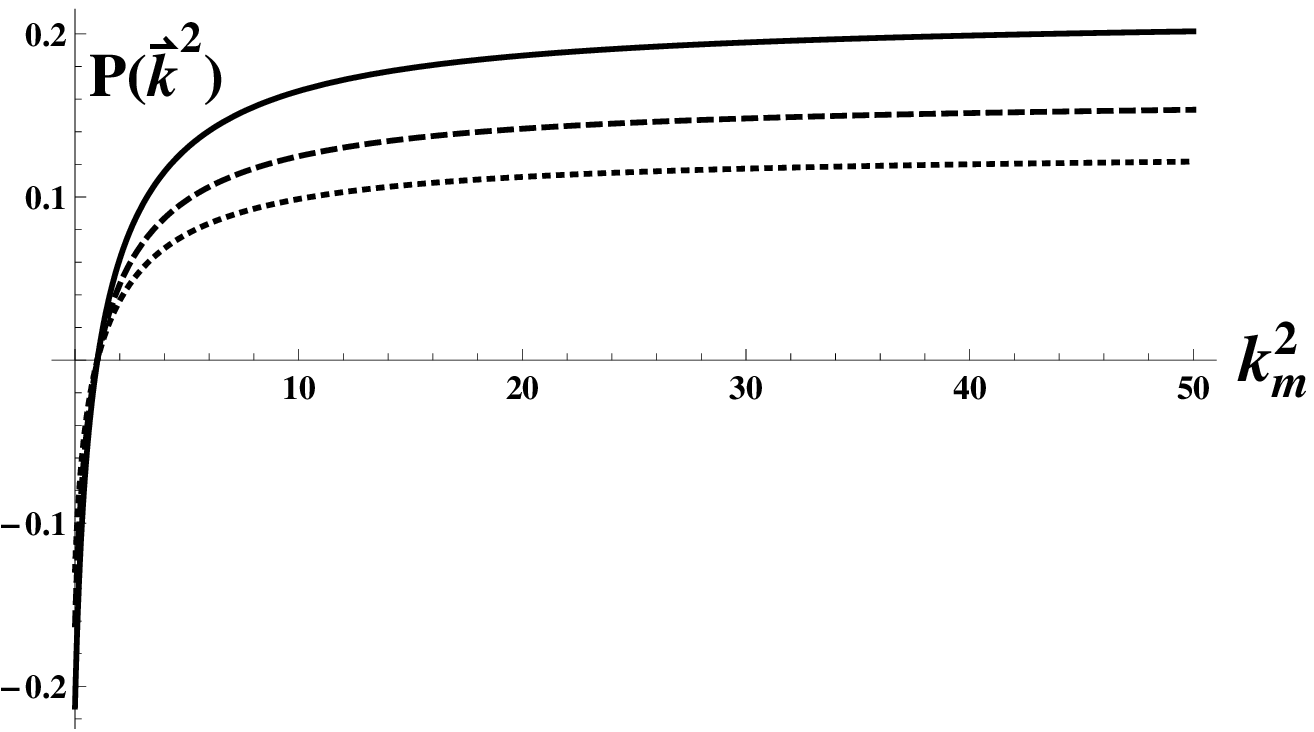} \\
\end{center}
\emph{\textbf{Fig.8.}} {\emph{Differential cross section as
defined by Eq.$\,$(32) and the respective polarization, that is
the ratio of the right hand side of Eq.$\,$(43) at $\xi_2=1$ to
the cross section (32), at $s=100\,MeV^2$ (solid curves),
$s=300\,MeV^2$ (dashed curves)
and $s=1\,GeV^2$ (dotted curves). }}\\
\end{minipage}
These relations distinguish from the corresponding ones but with
4-vector $S$ instead of $\tilde S$ (see formulas on page 15 after
Eq.~(35)). Just this distinction is a source of different forms of
the polarization dependent parts of the differential cross
sections. We take attention also that in accordance with Eq.~(43)
our respective spectral distribution vanishes for both
contributions: leading logarithmic and constant ones, whereas in
Ref.~\cite{BKGP02} the logarithmic contribution is non-zero
(Eq.~(16)).

Unpolarized cross section, within adopted accuracy, is symmetric
relative change $\beta \rightleftarrows (1-\beta).$ With our
choice of the 4-vector $S$ the created electron polarization is
antisymmetric if the recoil (or scattered) electron is recorded.
Otherwise, there are non-logarithmic contributions which do not
posses definite symmetry under this change (see Eqs.~(42), (45)).

The accuracy of our calculations is restricted by neglected terms
of the order of $m^2/s$ and by the radiative corrections. The
first ones can be essential near the boundaries of the electron
spectrum \cite{AAK00}. Therefore our calculations are valid for
the region $0.1<\beta< 0.9.$ As to the radiative corrections, they
violate the above mentioned symmetries on the percent level in
this region of the electron energies, at least for unpolarized
events, due to possibility of the hard photon emission
\cite{VKM74}.

\vspace{0.5cm}

\begin{minipage}{160 mm}
\begin{center}
\includegraphics[width=0.45\textwidth]{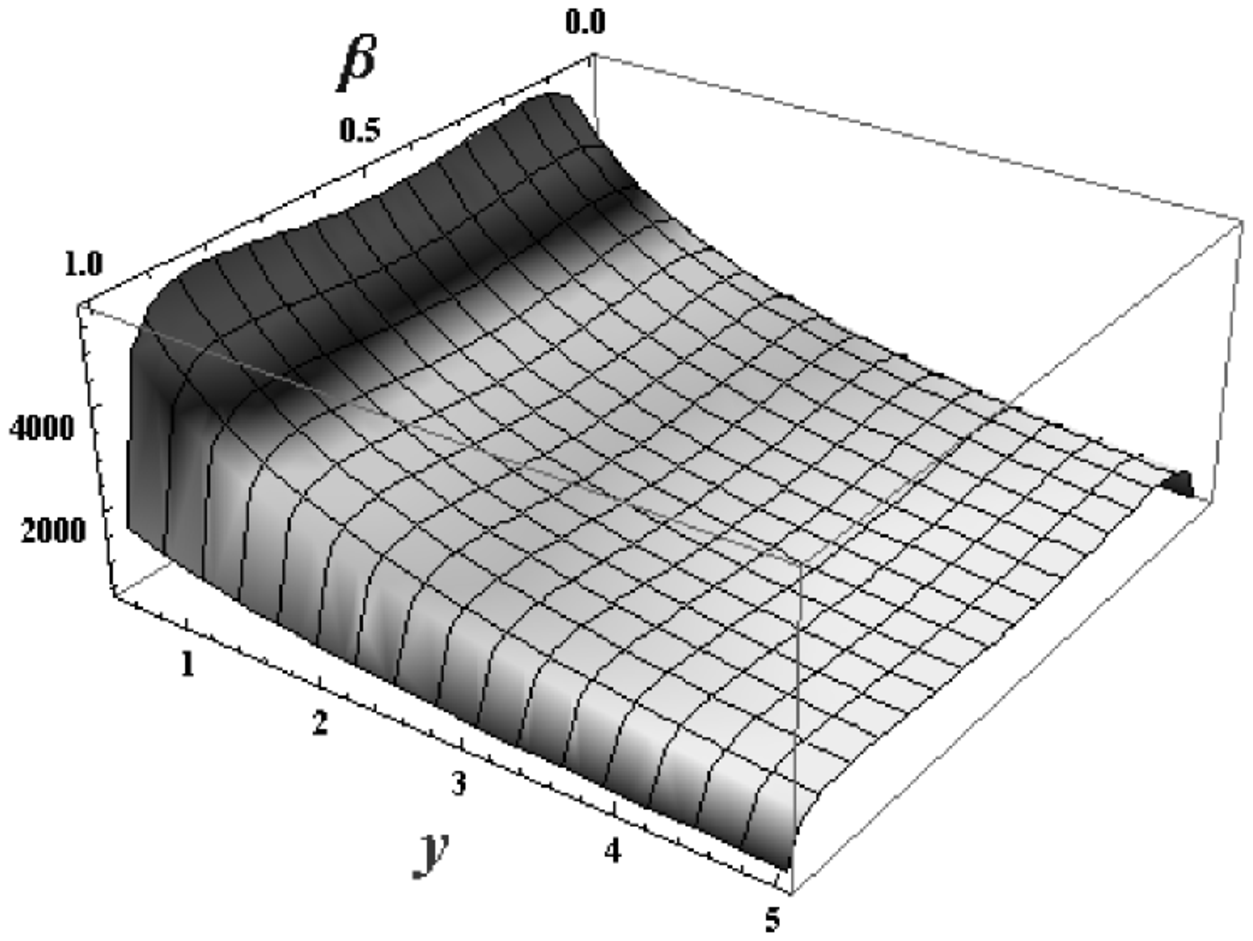}
 \hspace{0.4cm}
\includegraphics[width=0.45\textwidth]{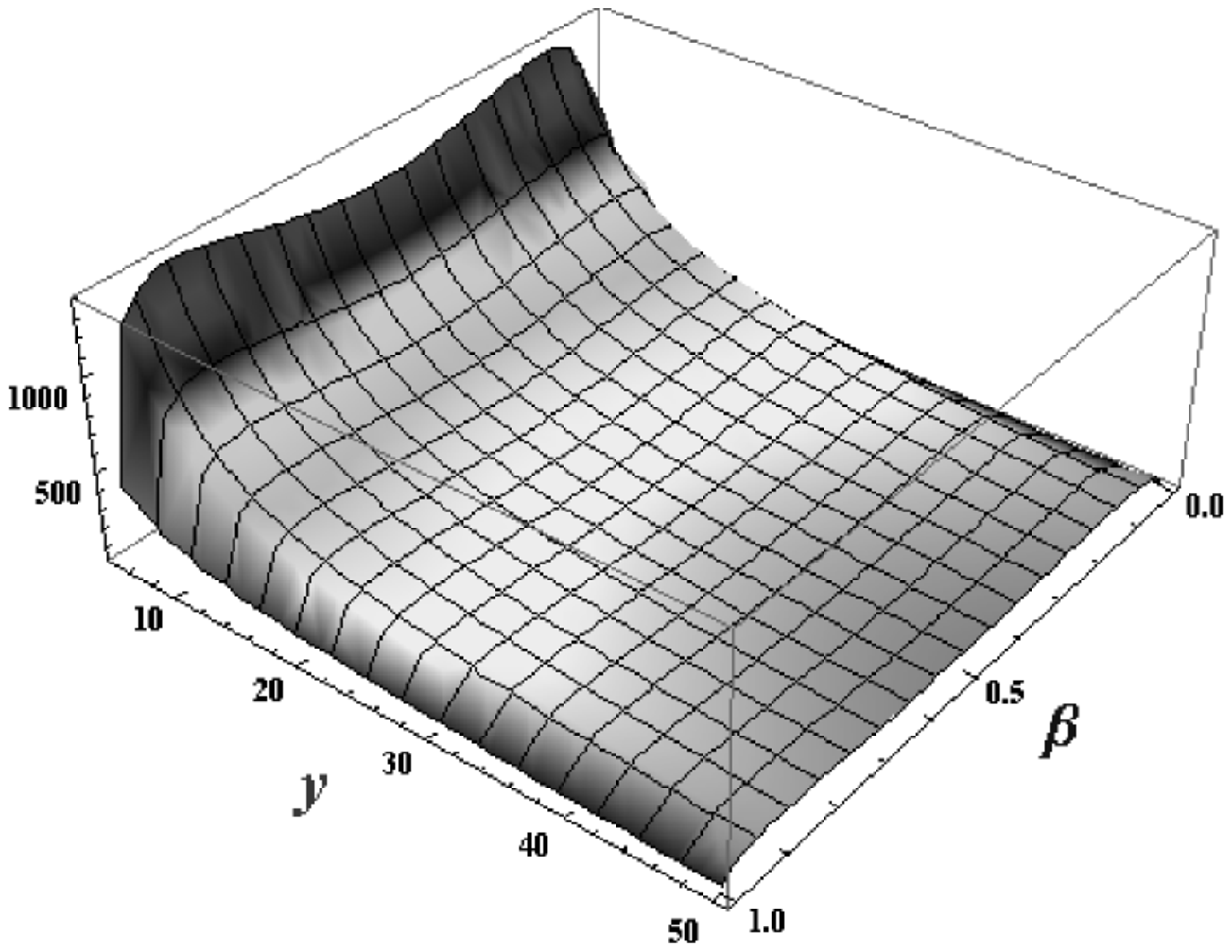}
\vspace{0.4cm}
\includegraphics[width=0.45\textwidth]{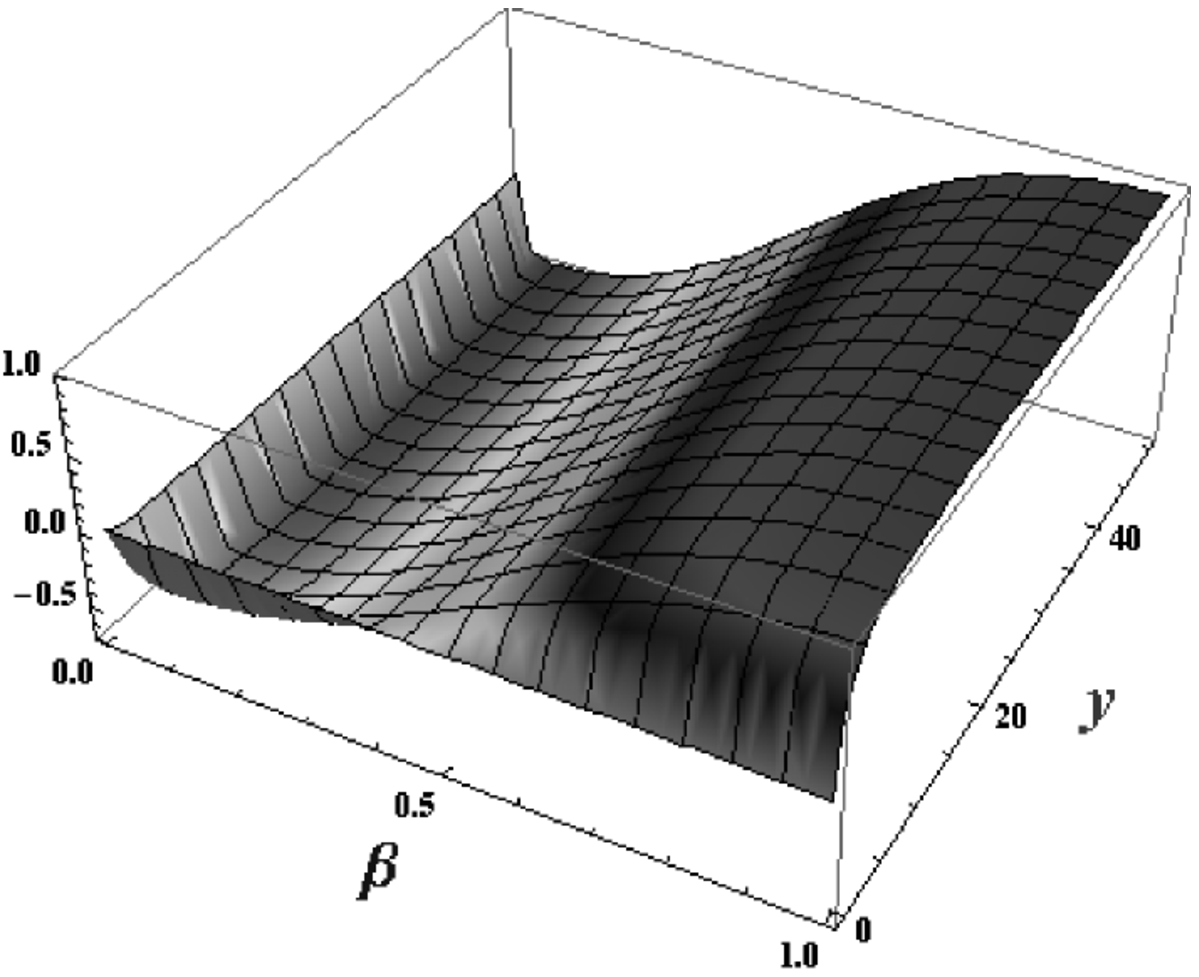} \\
\end{center}
\emph{\textbf{Fig.9.}} {\emph{Differential cross section (two
upper figures) and the created electron polarization (Eq. (45)) as
a function of the energy fraction $\beta$ and parameter $y$ at
$s=300\,MeV^2.$ }}\\
\end{minipage}

\section{Conclusion}

The process of the $e^+e^-$-pair production in the scattering of
the circularly polarized photon beam on the electrons leads to the
origin of the polarization of the produced electron and positron.
At high energy of the photon beam this effect can be used both for
the production of the high-energy polarized electron (and
positron) (see Ref. \cite{BKGP02}) and for the measurement of the
photon circular polarization degree since the differential cross
section and polarization transfer coefficient do not decrease with
the photon energy growth. The main contribution to these physical
quantities is caused by the events with small momentum transfer
squared ($|q^2|/s\ll 1$) when $e^+e^-$-pair carries away all
photon energy. This contribution is determined by the Borselino
diagrams (Fig. 1).

In our paper this contribution has been calculated for different
distribution of the final particles using the technique of the
Sudakov variables. We considered two essentially different
physical situations. The first one is concerned with the detection
not only produced electron but also the scattered (recoil)
electron. That kind of detection is quite possible since the final
electrons belong to different (non-overlapped) phase space
regions. The results of our numerical calculations are presented
in Figs. 3-6 for the case when minimal transverse transfer
momentum is of the order of the electron mass ($|{\vec
q}~^2|_{min}\approx m^2$).

The typical differential cross sections turn out to be of the
order of 1 mb and the polarization transfer coefficients are of
the order of 1 and antisymmetrical relative the change $\beta\to
(1-\beta ).$  Our calculations imply the integration over total
interval of the electron azimuthal angles. In principle, they can
be done for any detector geometry since the differential cross
section (the formulas (19) and (36)) is easy to integrate
numerically.

The results of calculations in case when the scattered electron is
not detected are presented in Figs. 7-9. In these calculations the
contributions of all events with $|{\vec q}~^2|\ge 0$ are added
together. The differential cross section (formulas (31) and (42))
acquire the contribution which growth logarithmically with the
energy. At the cost of this the cross sections turn out to be
somewhat larger than in the first case. The polarization transfer
coefficient is also of the order of 1 if the electron energy is
measured but it is essentially smaller if the integration over the
energy is done in all range of values (Fig. 8). It is important to
note that such experimental set up is possible in the interaction
of photons with the electron beam since during the interaction of
the photons with a matter  the scattering on the atomic electrons
with the $e^+e^-$-pair production (without recoil electron
detection) will be only the background process relative the
Bethe-Heitler process.

Authors thank A. Glamzdin for helpful discussion.

\end{document}